\documentclass{aastex62}

\usepackage{amsmath,amstext}
\usepackage[T1]{fontenc}
\usepackage[figure,figure*]{hypcap}
\usepackage{float}
\usepackage{wasysym}
\usepackage[normalem]{ulem}
\usepackage{enumitem,kantlipsum}

\graphicspath{{./}{}}
\received{May 20, 2020}
\revised{\today}
\submitjournal{ApJ}

\shortauthors{Perez et al.}

\def\lya{Ly$\alpha$}

\begin{document}

\title{Void Probability Function of Simulated Surveys of high-redshift Lyman-Alpha Emitters}

\correspondingauthor{Lucia A. Perez}
\email{lucia.perez@asu.edu}

\author[0000-0002-8449-1956]{Lucia A. Perez}
\affiliation{Arizona State University \\
School of Earth and Space Exploration \\
781 Terrace Mall \\
Tempe, AZ 85287, USA}

\author[0000-0002-9226-5350]{Sangeeta Malhotra}
\affiliation{NASA Goddard Space Flight Center \\
8800 Greenbelt Road \\
Greenbelt, MD 20771, USA}

\author[0000-0002-1501-454X]{James E. Rhoads}
\affiliation{NASA Goddard Space Flight Center \\
8800 Greenbelt Road \\
Greenbelt, MD 20771, USA}

\author[0000-0001-8514-7105]{Vithal Tilvi}
\affiliation{Arizona State University \\
School of Earth and Space Exploration \\
781 Terrace Mall \\
Tempe, AZ 85287, USA}

\begin{abstract}

We calculate the void probability function (VPF) in simulations of Lyman-$\alpha$ emitters (LAEs) across a wide redshift range ($z=3.1,\ 4.5,\ 5.7,\ 6.6$). The VPF measures the zero-point correlation function (i.e. places devoid of galaxies) and naturally connects to higher order correlation functions while being computationally simple to calculate. We explore the Poissonian and systematic errors on the VPF, specify its accuracy as a function of average source density and the volume probed, and provide the appropriate size scales to measure the VPF. At small radii the accuracy of the VPF is limited by galaxy density, while at large radii the VPF is limited by the number of independent volumes probed. We also offer guidelines for understanding and quantifying the error in the VPF. We approximate the error in the VPF by using independent sub-volumes of the catalogs, after finding that jackknife statistics underestimate the uncertainty.
We use the VPF to probe the strength of higher order correlation functions by measuring and examining the hierarchical scaling between the correlation functions using count-in-cells. 
The negative binomial model (NBM) has been shown to best describe the scaling between the two point correlation function and VPF for low-redshift galaxy observations. We further test the fit of the NBM by directly deriving the volume averaged two-point correlation function from the VPF and vice versa. We find the NBM best describes the $z=3.1, 4.5, 5.7$ simulated LAEs, with a 1$\sigma$ deviation from the model in the $z=6.6$ catalog. 
This suggests that LAEs show higher order clustering terms similar to those of normal low redshift galaxies. 

\end{abstract}

\section{Introduction} \label{sec:intro}

Galaxies are clustered in space according to the clustering of dark matter halos and the bias introduced by the baryonic physics involved in galaxy formation, both of which evolve over cosmic time (\citealt{Bernardeau2002,Benson2010,CoilReview2013}). The clustering of galaxies is is commonly studied with the two-point correlation function, which describes the probability of finding a galaxy within a given distance of another. The clustering can then be used to measure the power spectrum and constrain cosmological parameters. In observations, the two-point correlation function is used to measure the correlation length, which allows the comparison of clustering across galaxy samples and how galaxy clustering traces that of the dark matter halos. 
The clustering of galaxies not only depends on their relationships to their host dark matter halos and the growth of structure under gravity, but also is related in complicated ways to galaxy properties (like color and luminosity) and their local environment (eg. \citealt{Zehavi2002,Croton2005,Cooper2007, Skibba2014}).
Additionally, the volume-averaged correlation functions can be compared to perturbation theory and theoretical gravitational statistics \citep{PeeblesLSStextbook}, and are measured using the statistical moments of the galaxy count-in-cells distribution. While the standard two-point correlation function is commonly used in the literature, it  does not fully capture details of structure such as filaments and voids \citep{Maurogordato1987} and does not give information about higher order clustering correlations. 

The void probability function (VPF) is a less common clustering measurement that describes the probability that a sphere of a given size will contain no galaxies. Sometimes called the `zero-point' correlation function (the average distance where no galaxies exist), the VPF ties theoretically to the higher order correlation functions \citep{White1979}. The count-in-cells statistic includes the VPF and information about several averaged correlation function moments, and appears to follow a predictable pattern of scaling due to gravity. The VPF and two-point and higher order correlation functions are theoretically connected through this `hierarchical scaling', the idea that the first galaxies to form trace the first structures to collapse, following the evolution of Gaussian density fluctuations due to gravitational instabilities \citep{CoilReview2013}. In this framework, the VPF can connect to the higher order correlation functions under an analytic model for the scaling coefficients. The hierarchical scaling between the count-in-cells measured VPF and volume-averaged correlation functions has been found to follow the negative binomial model (NBM) in $z < 1$ galaxy surveys (e.g., \citealt{Croton2004b, Conroy2005, Tinker2008}) and in simulations (e.g., \citealt{Andrew2013}). \citet{Conroy2005} deduce and \citet{Fry2013} confirm that the NBM of clustering scaling is a feature of the underlying dark matter halos. Additionally, the VPF alone has been used to constrain cosmological parameters of large scale structure (ex. \citealt{Fry1986, Otto1986, Maurogordato1987, Fry1988, Little&Weinberg1994}). For example, the VPF was found to be sensitive to Halo Occupancy Distribution (HOD) models with different minimum host masses where the two-point angular correlation function was not (\citealt{Tinker2006, Berl&Weinb2002}) and was able to discern the influence of galaxy assembly bias on HOD modeling (\citealt{WalshTinker2019, BeltzMohrmann2020}).

To study the evolution of clustering across cosmic time, it is important to have a well-defined sample of galaxies that can be imaged over various redshifts. Lyman-$\alpha$ emitters (LAEs) have been observed in large numbers at redshifts $2 < z < 7$ through narrowband surveys (\citealt{Rhoads2000, Taniguchi2005, Shimasaku2006, Matthee2014, Santos2016, Zheng2016, Ouchi2018, Sobral2018a}). Lyman-$\alpha$ is the strongest emission signature of the earliest galaxies, likely coming from active star formation (\citealt{Partridge+Peebles1967, Malhotra+Rhoads2002, Atek2014}). Additionally, because many LAE properties do not seem to change significantly over time \citep{Malhotra2012}, they are excellent probes of the evolution of large scale structure and the process of reionization, whether by examining changes to their luminosity functions or clustering signatures brought about by neutral hydrogen attenuating the emission (\citealt{Malhotra+Rhoads2004, Mesinger+Furlanetto2008, Tilvi2014}). Understanding the clustering properties of LAEs and how they have evolved improves our understanding of galaxy evolution, the process and pace of reionization, and how galaxies are tied to their dark matter halos and environments. To date, several works have investigated the clustering properties of LAEs at various redshifts between $z \approx 2 - 7$ with mostly the angular two-point correlation function (e.g. \citealt{Ouchi2003, Shimasaku2004, Kashikawa2006, Gawiser2007, Kovac2007, Murayama2007,  Ouchi2008, Shioya2009, Ouchi2010, Guaita2010, Kusakabe2018b, SobacchiMesinger2015, Bielby2016, Hao2018, Ouchi2018, Khostovan2018b, Hong2019}). However, only \citet{Palunas2004} (34 LAEs at $z=2.38$), \citet{Kashikawa2006} (58 LAEs at $z=6.5$), \citet{McQuinn2007} (200 simulated LAEs at z=7.5 in partially neutral IGM) have used the VPF to measure the clustering of LAEs in the literature. 

As seen in this sampling of the greater field of galaxy clustering research, the VPF has been used in applications that do not often overlap or inform each other. The VPF has spanned diverse applications, whether used to detect slight signals of clustering for very small galaxy samples (eg. LAEs in \citealt{Palunas2004, Kashikawa2006}); to test hierarchical scaling in large spectroscopic samples of low redshift galaxies (eg. \citealt{Conroy2005,Croton2004b}); or as an additional constraint for theoretical models (eg. \citealt{Tinker2006, Berl&Weinb2002, WalshTinker2019,BeltzMohrmann2020}). In this work, we seek to clarify when and how to best use it in some of these applications. On what distance scales is a VPF measurement reliable? What might be the most honest way to approximate its uncertainty? How informative is the hierarchical scaling application of the VPF with smaller samples of galaxies? 
Can the VPF give information about higher-order clustering for high redshift starburst galaxies?
In local galaxies, the scaling behavior of higher order correlation functions has helped inform the influence of the local environment on galaxy clustering \citep{Baugh2004}, which is to be expected as galaxy properties are influenced by their environment (eg. \citealt{Croton2005}) and galaxy clustering and higher order correlation terms change with those same properties (eg. \citealt{Croton2004a,Conroy2005}). Can the VPF give similar insight for smaller samples of high redshift galaxies? Finally, how can these insights into the VPF inform future clustering studies within these frameworks and others?

In this work, we analyze simulations of LAE-inhabited dark matter halos across a wide redshift range \citep{Tilvi2009} to test the limits of the VPF. We measure the hierarchical scaling of the LAEs with count-in-cells, and test the fit by moving between the volume-averaged two-point correlation function and VPF in both directions. Our VPF measurements of the simulated LAE catalogs of \citet{Tilvi2009} can serve as comparisons to future clustering studies both in their measurements and how our uncertainty is calculated. We confirm that higher redshift LAEs are more clustered and their hierarchical scaling follows the negative binomial model to at least $z=6$.
This could indicate that, under the T09 model, Ly-$\alpha$ emission does not show
strong higher order clustering, though the scale of our simulations might not be enough to detect this signal.
Finally, the good agreement between the VPF and other clustering measurements we find shows that the VPF could replace the count-in-cells method in contexts that do not specifically probe hierarchical scaling (such as \citealt{Jensen2014} do to detect the influence of inhomogeneous reionization on the clustering of LAEs), while carrying through more intuitive errors that reflect the properties of the galaxy observations. 

This paper is organized as follows. In \textsection \ref{sec:TilviPaperInfo}, we give the relevant details of the \citet{Tilvi2009} LAE simulations. In \textsection \ref{sec:VPFGeneral}, we define the VPF, explore its uncertainty and error, and offer guidelines for its use. In \textsection \ref{sec:2ptCFs_Hierarchy}, we measure the volume-averaged and standard two-point correlation function, contextualize hierarchical scaling, and motivate the use of the negative binomial model. In \textsection \ref{sec:FitTestNBM}, we examine the hierarchical scaling behavior of our simulated LAEs. We measure the hierarchical scaling of our simulated LAEs with count-in-cells to show the negative binomial model is the best fit. We test the negative binomial model against our catalogs by deriving a VPF curve from the volume-averaged correlation function, as did \citealt{Conroy2005} (hereafter C05) and \citealt{Croton2004b} (hereafter Cr04). Then, we invert the transformation to derive the correlation function for these catalogs from their VPFs as an additional test of the negative binomial model. We give our main conclusions and final remarks in in \textsection \ref{sec:Conclusion}.

\section{Simulated Lyman-$\alpha$ Emitter Catalogs} \label{sec:TilviPaperInfo}

\citealt{Tilvi2009} (hereafter T09) presented a simple model of populating dark matter halos with Lyman-$\alpha$ emitters using large cosmological simulations. Here we briefly summarize relevant details and results. Their model assumes that \lya\ luminosity is proportional to the star formation rate (SFR), which is directly related to the mass accreted onto the host dark matter halo in the last 30 Myrs. 

T09 generated catalogs of dark matter halos using the $N$-body $\Lambda$CDM cosmological simulation GADGET2 \citep{GADGET2_2005} with initial conditions from second-order Lagrangian perturbation theory (\citealt{LagrangePerturb_Crocce, LagrangePerturb_ThnC}). The simulations contained $(1024)^3$ dark matter (DM) particles of mass $2.7 \times 10^7\ M_{\astrosun}\ h^{-1}$ in a volume of (102 cMpc)$^3$ or (73 $h^{-1}$ cMpc)$^3$. Using a friends-of-friends halo finder \citep{FoFhalofinder}, they tracked DM halos with at least 100 particles. The simulation was run from $z \approx 10$ to $z \approx 3$ while tracking the positions and masses of halos.

The DM halos with positive accretion rates are given a Lyman-$\alpha$ line luminosity using this model with star formation rate (SFR) and star formation efficiency (SFE):

\begin{equation}
    L_{\ \text{Ly}\alpha}\ =\ 1 \times 10^{42}\ \frac{\text{SFR}}{M_{\astrosun} \text{yr}^{-1}}\ \text{erg s}^{-1}\ ;\ 
    \text{SFR = SFE}\ \Big( \frac{\Delta M_{b}}{t_{\text{Ly}\alpha}} \Big)\ =\ \text{SFE} \Big( \frac{\Omega_{b}}{\Omega_{\text{DM}}} \Big) \Big( \frac{\Delta M_{\text{DM}}}{t_{\text{Ly}\alpha}} \Big)\
    \label{eq:TilviLAEmodel}
\end{equation}

Here, $\Delta M_{b}$ is the amount of baryonic mass accreted by the DM halos, and $t_{\text{Ly}\alpha} = 30$ Myr is the short timescale over which this accreted mass is converted into new stars in the model. T09 chose $t_{\text{Ly}\alpha} = 30$ Myr based broadly on the ages of stars in observed LAEs, the lifetime of OB associations, and the dynamical time that the size of LAEs predicts (e.g. \citealt{Finkelstein2007}; \citealt{Pirzkal2007}; \citealt{Finkelstein2008}). The accreted baryonic mass is derived from the universal ratio of baryonic to DM densities ($\Omega_{b},\ \Omega_{\text{DM}}$), and the mass accreted onto a DM halo at each step of the simulation ($\Delta M_{\text{DM}}$). The model assumed that the escape fraction of \lya\ is 1 and that of Lyman continuum photons is 0. In observations, the \lya\ escape fraction of LAEs has been measured at about 10 to 50 percent (\citealt{Nakajima2012}; \citealt{Matthee2016}; \citealt{Sobral2018a}), and possibly increasing with redshift and with complicated relationships to \lya\ equivalent width (\citealt{Sobral2017}; \citealt{Harikane2018}; \citealt{Oyarzun2017}; \citealt{Trainor2019}).

\begin{table*}[t]
	\begin{center}
    \caption{Relevant details from \citet{Tilvi2009} for clustering measurements. The column headers stand for: redshift (at the center of the simulation box); the \lya\ line luminosity cut applied to match the catalog to the corresponding comparison observation(s); the total number of simulated LAEs in the catalog; the number of LAEs after the luminosity cut; the surface density of galaxies after the luminosity cut in arcmin$^{-2}$, for the whole volume and two halves when cut evenly in redshift space; the volume density of galaxies after the luminosity cut in cMpc$^{-3}$, with each side of the volume measuring 102 cMpc; and the observations of LAEs to which the catalogs were designed to compare. The `front' is the area with $z$-coordinates from 0 to 51 cMpc, and the `back' is $z$-coordinates from 51 to 102 cMpc. The $z=4.5$ catalog has had the `back' left quarter removed due to a previously undiscovered output error in the simulation.}   
	\begin{tabular}{ p{.6cm}p{1.5cm}p{1cm}p{1cm}p{5.5cm}p{2.25cm}p{3.5cm} } 
	\hline \hline
    \textit{z} & L$_{\text{cut}}$, $10^{42}$ erg s$^{-1}$ & N$_{\text{total}}$ & N$_{\text{cut}}$ & Surface Density $\Sigma_{\text{arcmin}^{-2}}$& Volume Density $\mathcal{N}_{\text{Mpc}^{-3}}$ &  Comparison observations\\ 
    \hline
		3.1 & 1.995 & 62,364 & 1,145 & Whole $(102\ $ cMpc$)^3$: 0.4029 & 1.079 $\times\ 10^{-3}$ & \citet{Gawiser2007} $\&$\\
		&   &   &   & Front $102 \times 102 \times 51$ cMpc$^3$: 0.1717  &  & \citet{Khostovan2018b} \\
		&   &   &   & Back $102 \times 102 \times 51$ cMpc$^3$: 0.2312  & & \\
        
        4.5 & 1.673 & 64,868 & 1,211 & Whole $(0.75\ \times\ (102 $ cMpc$)^3)$: 0.5765 & 1.521 $\times\ 10^{-3}$ & \citet{Kovac2007} \\
        &   &   &   & Front $102 \times 102 \times 51$ cMpc$^3$:  0.3842 & & \citet{Ouchi2003}\\
        &   &   &   & Back $51 \times 102 \times 51$ cMpc$^3$: 0.3847  & & \\
        
        5.7	& 3.068 & 79,429 & 539 & Whole $(102\ $ cMpc$)^3$: 0.3014 & 5.08 $\times\ 10^{-4}$& \citet{Ouchi2010} $\&$ \\
        &   &   &   & Front $102 \times 102 \times 51$ cMpc$^3$: 0.1197  & & \citet{Ouchi2008}\\
		&   &   &   & Back $102 \times 102 \times 51$ cMpc$^3$: 0.1817  & & \\
        6.6	& 3.068 & 79,783 & 355 & Whole $(102\ $ cMpc$)^3$: 0.2171 & 3.35 $\times\ 10^{-4}$ &  \citet{Ouchi2010}\\
        &   &   &   & Front $102 \times 102 \times 51$ cMpc$^3$: 0.0960  & & \\
		&   &   &   & Back $102 \times 102 \times 51$ cMpc$^3$: 0.1211  & & \\
    \hline \hline	
    \\
	\end{tabular}

	\end{center}
    \label{table:TilviInfo}
\end{table*}

In the simple model of T09, LAEs fundamentally act as tracers of DM halo buildup and the accompanying cold gas accretion. T09 apply an analytic description to populate dark matter halos with LAEs based only on the evolution of the $N$-body simulation. Progress from many angles has occurred in the field of LAE modeling since. Similar studies that worked primarily with smoothed particle hydrodynamic simulations and intrinsic galaxy properties include: \citealt{Nagamine2010} (who also consider the intrinsic star formation in LAEs, but do not consider dust enrichment or variance in the intergalactic medium transmission); \citealt{Dayal2009} and \citealt{Dayal2010} (whose model includes in the luminosity of stellar sources, cooling of HI, dust enrichment, and IGM transmission); and \citealt{Kobayashi2007} and \citealt{Kobayashi2010} (who apply the Mitaka semianalytical hierarchical clustering model for star formation from \citealt{NagashimaYoshii2004} to account for extinction from interstellar dust and outflow feedback). Many works have also included radiative transfer calculations, for example to prepare for epoch of reionization by expanding how their models account for the effect of neutral IGM on the \lya\ line (\citealt{McQuinn2007}; \citealt{Iliev2008}; \citealt{Zheng2010}; \citealt{Dayal2011} \citealt{Jensen2013}; \citealt{Kakiichi2016}; \citealt{Inoue2018}; \citealt{Gangolli2020}; and more.) 

The simple model of T09 fits the single parameter of star formation efficiency to find good agreement with LAE observations at redshifts 3.1, 4.5, 5.7, and 6.6. The \lya\ luminosity function of the $z=3.1$ simulation is used to set the SFE by comparing with the LF of the \citet{Gronwall2007} LAE observations at the same redshift. With all parameters defined or derived in the model, T09 match the \lya\ line luminosity limits of surveys of LAEs at the same redshifts to test the strength their model. They reproduce the \lya\ luminosity functions, star formation rates, approximate halo masses, and duty cycles of comparable LAE surveys and observations with a simple adjustable parameter for all redshifts. Finally, their derived correlation lengths track closely with those observed for LAEs at similar redshifts. These simulations are therefore ideal for this exploration of the VPF for LAEs at varying redshifts.

For this work, we use the output catalogs that give every LAE's position in the $x-y-z$ space (between 0 and 73 $h^{-1}$ cMPc) and their Lyman-$\alpha$ line luminosity. For the z=4.5 catalog, we exclude the back left quadrant of $x <$ 36 $h^{-1}$ cMPc and $z > $ 36 $h^{-1}$ cMPc (the top and bottom `back' left (51 cMpc)$^3$ sub-volumes) because the simulation save file for this redshift was apparently truncated, omitting most DM halos (and therefore LAEs) from this subvolume. The rest of the volume behaved as expected, so we exclude this artificially empty region in our analysis. Accounting for this improves the agreement of the z=4.5 simulated LAEs with the comparison observations of \citealt{Kovac2007} (hereafter K07) and \citealt{Ouchi2003} (hereafter O03), and explains the slight inflation of the $z=4.5$ correlation length in T09.

When plotting our VPF measurements, we convert the transverse comoving positions to transverse comoving megaparsec (hereafter cMpc) using T09's $h=0.716$, so that the positions range from 0 and 102 cMPc. When calculating all correlation lengths, we revert to units of $h^{-1}$ cMpc to directly compare with earlier measurements of observed LAEs, whose angular separations were converted into physical distances assuming $h=0.7$ (from $H_0 \equiv$ 100 $h$ km s$^{-1}$ Mpc$^{-1}$ = 70 km s$^{-1}$ Mpc$^{-1}$). 

Table \ref{table:TilviInfo} provides relevant details of the catalogs for this work. The luminosity cuts follow the  survey limits for observations that each catalog mimicked. We confirm the corresponding Lyman-$\alpha$ line luminosity for the $z= 4.5$ K07 observation and confirm the luminosities that \citet{Ouchi2010} found for the observations at $z = 3.1,\ 5.7,$ and $6.6$ by using the relevant limiting magnitude and narrowband filter details for each observation. We also compare to the LAE sample of \citet{Khostovan2018b} at $z=3.10$ in the SA22 field using archival narrowband imaging that closely matches the \lya\ luminosity limit and volume density of our $z=3.1$ simulation catalog.

The simulated LAEs of T09 successfully recreated observed luminosity functions, equivalent width distributions, age distributions, and duty cycles. Future \lya\ surveys can further test the \citet{Tilvi2009} model by a similar comparison of our VPF measurements and hierarchical scaling.

\section{Void Probability Function} \label{sec:VPFGeneral}

\subsection{VPF Theory and Algorithm} \label{subsec:VPFtheoryncode}
The Void Probability Function (VPF) is the probability that regions of a particular radius will have no galaxies within them. It contains information about all higher order correlations  (\citealt{White1979,Maurogordato1987}), and is the `zero-point' volume-averaged correlation function. The VPF (labeled $P_0$ for brevity) of a galaxy sample with mean density $n$ is defined by the hierarchy of all $p$-point reduced  correlation functions $w_p$ at a given volume $V$:

\begin{equation}
    P_0(V) = \exp \Big( \sum^{\infty}_{p=1} \frac{(-n)^p}{p!}  \int ... \int w_p(x_1,...,x_p)\ dV_1... dV_p \Big)
    \label{eq:VPFdefinition}
\end{equation}

The VPF is calculated by simply counting what fraction of randomly placed test spheres are empty as a function of radius. Our algorithm takes in galaxy locations as point sources on an $x-y$ (2D) plane or $x-y-z$ (3D) volume, and for each radius being tested, generates many random central positions within the area (that would not have the radius overlap the boundaries), and counts how many of the test spheres have no galaxies within the radius. 

When we generate the randomly placed test spheres, we place ten thousand points to guarantee we sample all true voids. We repeat the VPF measurement for all radii 50 times to account for the small variance that comes when choosing random points and guarantee that we minimize the error from not completely sampling our volumes (hereon out the \textit{sampling error}, which C05 used as the error on their VPF). Throughout this work, we plot errors in relative logarithmic space:

\begin{equation}
    Y=\log_{10}(P_0)\quad \rightarrow \quad \delta Y\ \approx 0.434 \times \frac{\delta(P_0)}{P_0}
\label{eq:LogError}
\end{equation}

Figure \ref{fig:VPF2Dn3Dallcats} displays the VPF calculations of our four catalogs as colored stars (3D) or colored circles and triangles (2D, `front' and `back' halves). The 2D VPF is the most easily implemented tool for observational surveys, since redshift space data are arduous to gather and introduce additional error, while the angular positions of objects in the sky are known with excellent precision. When there exist accurate redshift data, the three-dimensional VPF in redshift space can be explicitly connected to the correlation functions by assuming given hierarchical models (eg. C05). \textsection \ref{sec:FitTestNBM} details the connection between the three-dimensional VPF and volume-averaged three-dimensional two-point correlation function for these simulations. 

\subsection{Verifying our VPF algorithm}\label{subsec:CheckVPFcode}

We check the accuracy of our VPF calculation with randomly distributed points. The original equation for the VPF from \citet{White1979} gives its exact value for a Poisson distribution. This theoretical VPF curve of completely unclustered points depends only on the surface or volume density of the sample. With \textit{N} as the number of galaxies sought in the volume \textit{V}, and \textit{n} as the mean particle density, the probability to find \textit{N} galaxies in a volume of \textit{V} is:

\begin{equation}
P_N(V) = \frac{(nV)^N}{N!} e^{-nV}\ ;\quad \text{for} \quad  N\ =\ 0\ ,\ P_0(V) = e^{-nV} .
\label{eq:VPFrandomptsOG}
\end{equation}

We can rewrite this equation to use surface density $\Sigma$ or volume density $\mathcal{N}$ and the radius of the tested void to yield these expressions for the VPF of completely unclustered points:

\begin{equation}
\log_{10}(P_{0, \text{2D}}(r)) = \frac{-\Sigma \times (\pi R^2)}{\ln(10)}
\label{eq:VPFrandomptsSDens}
\end{equation}

\begin{equation}
\log_{10}(P_{0, \text{3D}}(r)) = \frac{-\mathcal{N} \times ((4/3)\pi R^3)}{\ln(10)}
\label{eq:VPFrandomptsVDens}
\end{equation}

Deviation from this VPF indicates a sample is clustered. We check our algorithm by generating randomly placed points at the same surface or volume density of the given catalog after the luminosity cuts; measuring their VPF; and finally comparing the curve to the theoretical predictions. We find excellent agreement between the theoretical curves and the measured VPF of randomly placed points, confirming our algorithm is accurately measuring the VPF. 

\begin{figure*}
  \centering
\includegraphics[width=.94\textwidth]{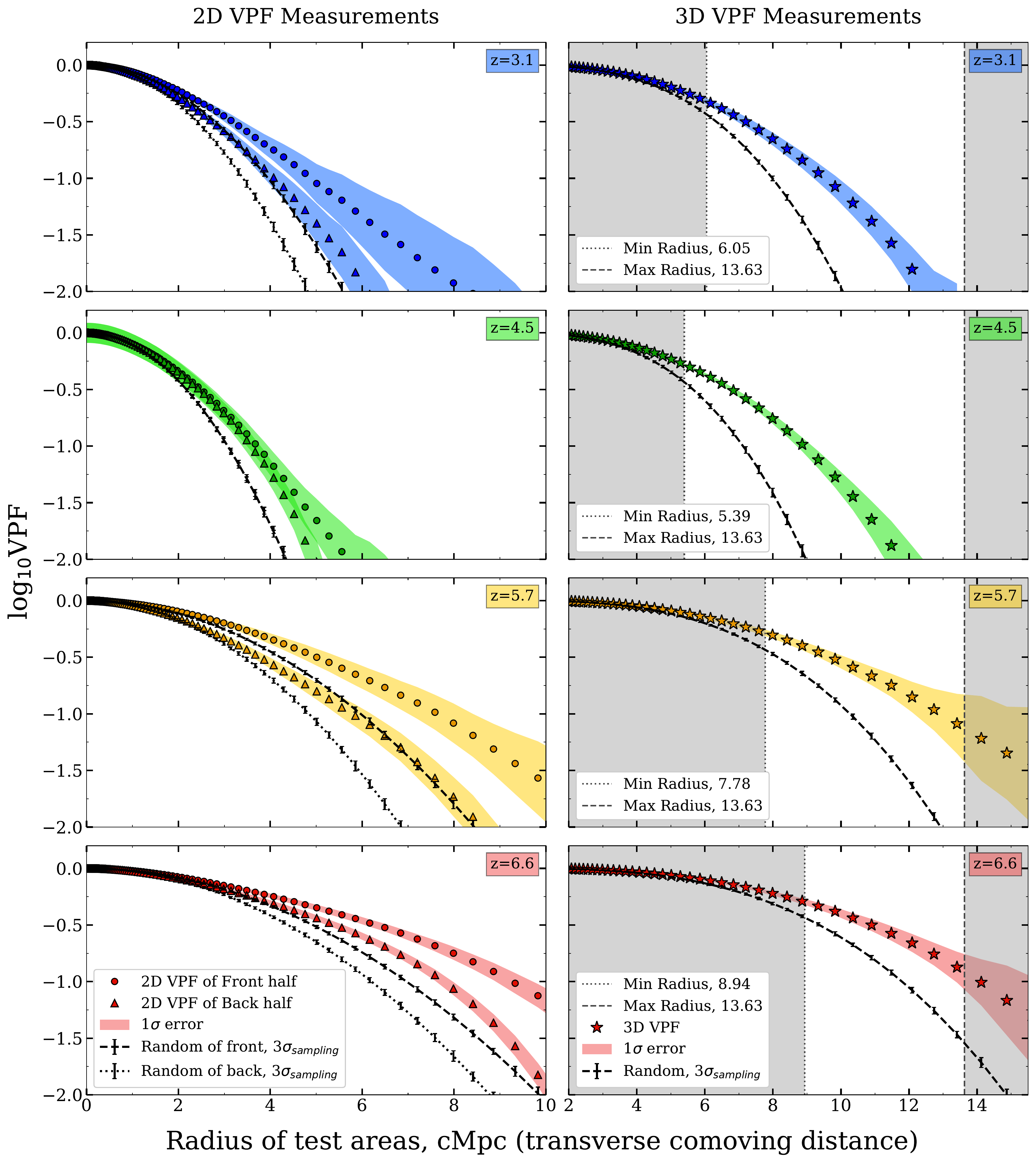}
   \caption{\textbf{\textit{Left: the measured 2D VPF.}} For the 2D VPF, we have split our volumes into `front' and `back' halves (split halfway in the $z$ direction) to match the thickness of the 8 independent cubical sub-volumes. The circles/triangles mark the 2D VPF of the front/back halves. The black dashed/dotted lines trace the 2D VPF of unclustered points at the same surface density of the front/back halves. \textbf{\textit{Right: the measured 3D VPF}}. For the 3D VPF, we have marked the theoretical radii where the VPF can be measured based on the number densities, volume, and our chosen precision of $\pm 10^{-2}$. The stars are the 3D VPF of the whole (102 cMpc)$^3$ volume (0.75 $\times$ (102 cMpc)$^3$ for the $z=4.5$ catalog), and the dashed black line is the 3D VPF of unclustered points at the same volume density. The error bars around the black lines correspond to 3 times the sampling error. The shaded colored regions correspond to $1\sigma$ standard error across the independent (51 cMpc)$^3$ sub-volumes' VPFs. All our catalogs' VPFs differ significantly from their random VPF curves, confirming that they are all clustered.} 
    \label{fig:VPF2Dn3Dallcats}
\end{figure*}

Figure \ref{fig:VPF2Dn3Dallcats} displays the theoretical VPF curves for the relevant density as dashed (3D, 2D `front' half) or dotted (2D `back' half) black lines. All our catalogs' VPF curves differ significantly from their random VPF predictions, confirming that they are all clustered. The `front' is the area with $z$-coordinates from 0 to 51 cMpc, and the `back' is $z$-coordinates from 51 to 102 cMpc. The error bars corresponding to the $3\sigma$ sampling error are plotted around these unclustered VPF curves. The minuscule sampling error confirms that the size of our our galaxy samples and test sphere arrays is large enough when compared to our full volume for a precise measurement.

\subsection{Fundamental Limits of the VPF} \label{subsec:TheoryErroronVPF}

Multiple factors contribute to the uncertainty in clustering measurements, some of which have been underestimated in past studies of the VPF. For the VPF, the most fundamental factors are the size and scale of the galaxy sample. 

To articulate these dominating constraints to a reliable VPF measurement, we imagine a distribution of $N$ galaxies in a volume of $V$ with number density $\mathcal{N}$, randomly distributed and with no correlation in their positions. The smallest radius to dependably probe the VPF is the volume inhabited by a single galaxy, or the average distance between two galaxies: 

\begin{equation}
    \mathcal{N} \frac{4 \pi}{3} R_{\text{min}}^3 = 1 \quad \rightarrow \quad R_{\text{min}} = \sqrt[3]{\frac{1}{\mathcal{N}} \frac{3}{4 \pi}}
\label{eq:RminTheor}
\end{equation}

The largest radius to dependably probe the VPF depends instead on the size of the volume and the desired precision of the VPF. The VPF is a fraction often reported as logarithms, so in order to measure $\log_{10}(P_0)$ to a given $-\alpha$ value and guarantee the level of precision of the VPF within $\pm 10^{-\alpha}$, the volume must accommodate $10^{\alpha}$ independent sub-volumes of the given radius. For example, if one of these $10^{\alpha}$ independent sub-volumes is empty, then the VPF is indeed $1/10^{\alpha}$. Therefore, the maximum radius to probe VPF is the radius of the independent sub-volumes that probe the desired sensitivity in the sample volume:

\begin{equation}
    V =  10^{\alpha} \frac{4 \pi}{3}  R_{\text{max}}^3 \quad \rightarrow \quad R_{\text{max}} = \sqrt[3]{\frac{V}{10^{\alpha}} \frac{3}{4 \pi}}
\label{eq:RmaxTheor}
\end{equation}

Combining these minimum and maximum radii measurements allows us to constrain the number of galaxies needed in a sample for the desired measurement. The dynamic range between the radii is then:

\begin{equation}
    d = \frac{R_{\text{max}}}{R_{\text{min}}} = \frac{ ((V/10^{\alpha}) (3/4\pi))^{1/3} }{ ((1/\mathcal{N}) (3/4\pi))^{1/3} } \quad \rightarrow \quad \sqrt[3]{\frac{\mathcal{N}V}{10^{\alpha}}}
\label{eq:DynRangeTheor}
\end{equation}

A specific dynamic range and VPF sensitivity will constrain a specific number density and volume ratio, and therefore the number of galaxies for the measurement:

\begin{equation}
    N_{\text{total}} = \mathcal{N} V = d^3 10^{\alpha}
\label{eq:nV_N_Theor}
\end{equation}

For example, with a dynamic range of 2 in radius (eg. 6 to 12 cMpc) and probing to $\log_{10}(P_0) = -2$, the sample must include at least 800 galaxies. To improve the measurement of the VPF at small scales, we must increase the number density. To improve the measurement at large scales, we must increase the volume.

These derivations rely on assuming a completely unclustered galaxy distribution, and are therefore very conservative guidelines. When galaxy samples display clustering, the average distance between galaxies is smaller and the limiting $R_{\text{min}}$ will decrease. For our samples, we probe to $\log_{10}(P_0) = -2$. All catalogs have the same underlying (102 cMpc)$^3$ volumes and same maximum radius of 13.63 cMpc.\footnote{Although we chose remove 25$\%$ of the volume in the $z=4.5$ simulation to account for a previously undiscovered output error, we maintain this maximum radius for consistency. This choice makes little difference in our final result.} Table \ref{table:TheorLimitsInfo} shows the minimum radii and dynamic ranges of our catalogs after applying the luminosity cuts.

\begin{table}
	\begin{center}
    \caption{Theoretical radii VPF limits for our catalogs, assuming $\alpha$=2 and $V$=(102 cMpc)$^3$ for a perfectly random distribution. Based on the number of galaxies left after the luminosity cut and the resulting number density of the sample, these are the minimum radii and largest dynamical range of radii over which the VPF can be measured.}   
	\begin{tabular}{ p{.6cm}p{1cm}p{2.5cm}p{2cm}p{1cm} } 
	\hline \hline
    \textit{z} & N$_{\text{cut}}$ & $\mathcal{N}_{\text{cMpc}^{-3}}$ & R$_{\text{min}}$, cMpc & $d$  \\ 
    \hline
		3.1 & 1,145 &  1.079 $\times\ 10^{-3}$ & 6.05  & 2.25 \\

        4.5 & 1,211 &  1.521 $\times\ 10^{-3}$ & 5.39  & 2.53 \\

        5.7 & 539   &  5.08 $\times\ 10^{-4}$  & 7.78  & 1.75 \\

        6.6 & 355   &  3.35 $\times\ 10^{-4}$ & 8.94   & 1.52 \\
    \hline \hline	
	\end{tabular}
	\end{center}
    \label{table:TheorLimitsInfo}
\end{table}

In Figure \ref{fig:VPF2Dn3Dallcats}, we indicate these radii limits for our 3D VPFs with grey shading. In \textsection \ref{subsec:TheoryErroronVPF}, we restrict ourselves to these radii when we convert our catalog's VPFs into correlation functions. The luminosity cuts we used so we could directly compare to observations decrease the number density and increase $R_{\text{min}}$ significantly. Upon examination of Figure \ref{fig:VPF2Dn3Dallcats}, the limiting radii in Table \ref{table:TheorLimitsInfo} correspond to when the random and VPF curves diverge significantly, and when the error begins to dominate.

Though we focus on the 3D VPF for the rest of this work, we can also derive these radii limits for the 2D VPF using the survey area $A$ and surface density $\Sigma$:

\begin{equation}
    \Sigma \times \pi R_{\text{min}}^2 = 1 \quad \rightarrow \quad R_{\text{min}} = \sqrt{\frac{1}{\Sigma \pi}}
\label{eq:RminTheor_2D}
\end{equation}

\begin{equation}
    A =  10^{\alpha} \times \pi  R_{\text{max}}^2 \quad \rightarrow \quad R_{\text{max}} = \sqrt{\frac{A}{\pi 10^{\alpha}}}
\label{eq:RmaxTheor_2D}
\end{equation}

\begin{equation}
    d = \frac{R_{\text{max}}}{R_{\text{min}}} =  \sqrt{\frac{\Sigma A}{10^{\alpha}}}
\label{eq:DynRangeTheor_2D}
\end{equation}

\begin{equation}
    N_{\text{total}} = \Sigma \times A = d^2 10^{\alpha}
\label{eq:nV_N_Theor_2D}
\end{equation}

\subsection{Traditional statistical estimators for error in the VPF} \label{subsec:PastErroronVPF}

The majority of past studies utilizing the VPF have approximated the error in their measurements by utilizing traditional methods of `internal' error estimates or creating mock catalogs and measurements (`external' estimates). `Internal' error estimates cut the entire sample into smaller sections and repeat the statistical measurement in different iterations. \citet{Norberg2009} give an excellent review and assessment of internal error estimation methods for correlation function clustering studies. They determine that the bootstrap method with oversampled sub-volumes agrees best with their external error estimation methods (galaxy formation models with fully known inputs). In the \textit{bootstrap} method, a resampling of the data set is made by randomly selecting some number of sub-volumes while allowing for replacement and repetition.

However, conclusions for the two-point correlation function error behavior will not exactly translate to the VPF. Past studies using the VPF observed galaxies often use the \textit{jackknife} method, where the volume is divided into subsections, each subsection is systematically left out the full volume for a measurement, and the mean and variance of the final distribution is taken to approximate the mean and variance of the original. For example, C05 exclusively use the jackknife technique to estimate their errors on their VPF. Cr04 quote 1$\sigma$ errors on their VPFs derived from the \textit{rms} scatter over their many generated mock catalogs, and verify that the error from jackknife technique is comparable. Finally, \citet{Khostovan2018a} found that the Poisson estimate of the correlation function's error is very consistent with the bootstrapping errors for samples with $10^{2-3}$ galaxies. 

We measure the variance with the traditional jackknife technique with the eight sub-volumes and find it underestimates the error compared to the subsample method we introduce in \textsection \ref{subsec:ErrorVPFcatalog} by at least a factor of 2 at all scales. \citet{YangSaslaw2011} suggested that the jackknife method of C05 underestimated the true range of the data's variability, as subsets of the data are not identically distributed. Therefore, we caution against relying solely on this technique to measure the error on the VPF measurement, while recognizing that more work remains to be done studying how the jackknife technique performs on the VPF of larger samples.

\subsection{Approximating error in the VPF with independent sub-volumes} \label{subsec:ErrorVPFcatalog}

We choose the subsample method to quantify our error in the most `independent' way possible. Like other gauges of clustering, individual measurements of the VPF are not independent, since clustering also takes place on scales much larger than most observations. In a practical sense, a single abnormally empty or overdense region affects the entire VPF curve for the whole volume. To mimic how truly independent measurements of the VPF within our radii limits would behave, we divide our catalog volumes into the largest equally sized independent subvolumes that they hold (72 cMpc h$^{-1}$ $\div$ 2 = 36 cMpc h$^{-1}$ or 51 cMpc per side), and use the spread in the subvolumes' VPF measurements to guide the error in our whole volume's VPF. Though such subdivisions of galaxy data are not truly independent, this method helps illuminate the inherent variance of the VPF at radii less than 15 Mpc.

In Figure \ref{fig:VPF2Dn3Dallcats}, we shade the subsample error around the VPFs of the `halves' (2D) and the whole volume (3D) in the corresponding colors.
For this method, we divide our simulated catalogs into eight non-overlapping independent sub-volumes of (51 cMpc)$^3$. (The $z=4.5$ catalog ends up with six non-overlapping independent sub-volumes, due to the apparent output error in the back left corner that was not discovered until this work.) We measure the VPF ten times for each of the sub-volumes with 5000 randomly dropped test voids every time. For an approximation of the error of the whole volumes' VPF at each radius, we measure the standard error in the population mean (the standard deviation of the 80 VPF measurements across all sub-volumes and iterations divided by the square root of 8):

\newcommand{\overbar}[1]{\mkern 1.5mu\overline{\mkern-1.5mu#1\mkern-1.5mu}\mkern 1.5mu}
\begin{equation}
       1\sigma\ \text{standard error}\ =\  \frac{\text{STD}(P_{0, \text{subvols}})}{\sqrt{N_{\text{subvols}}}} \quad \rightarrow \quad
       \frac{\sqrt{\overline{ \lvert P_{0,i} - \bar{P_0} \rvert^2 }}}{\sqrt{8}}
\label{eq:VPF_standardError} 
\end{equation}

\section{Hierarchical Scaling and the Correlation Functions} \label{sec:2ptCFs_Hierarchy}

\subsection{Measured Two-Point Correlation Functions $\&$ Correlation Lengths} \label{subsec:2ptcorrfunc}

\begin{figure*}
	\begin{center}
    \includegraphics[width=\textwidth]{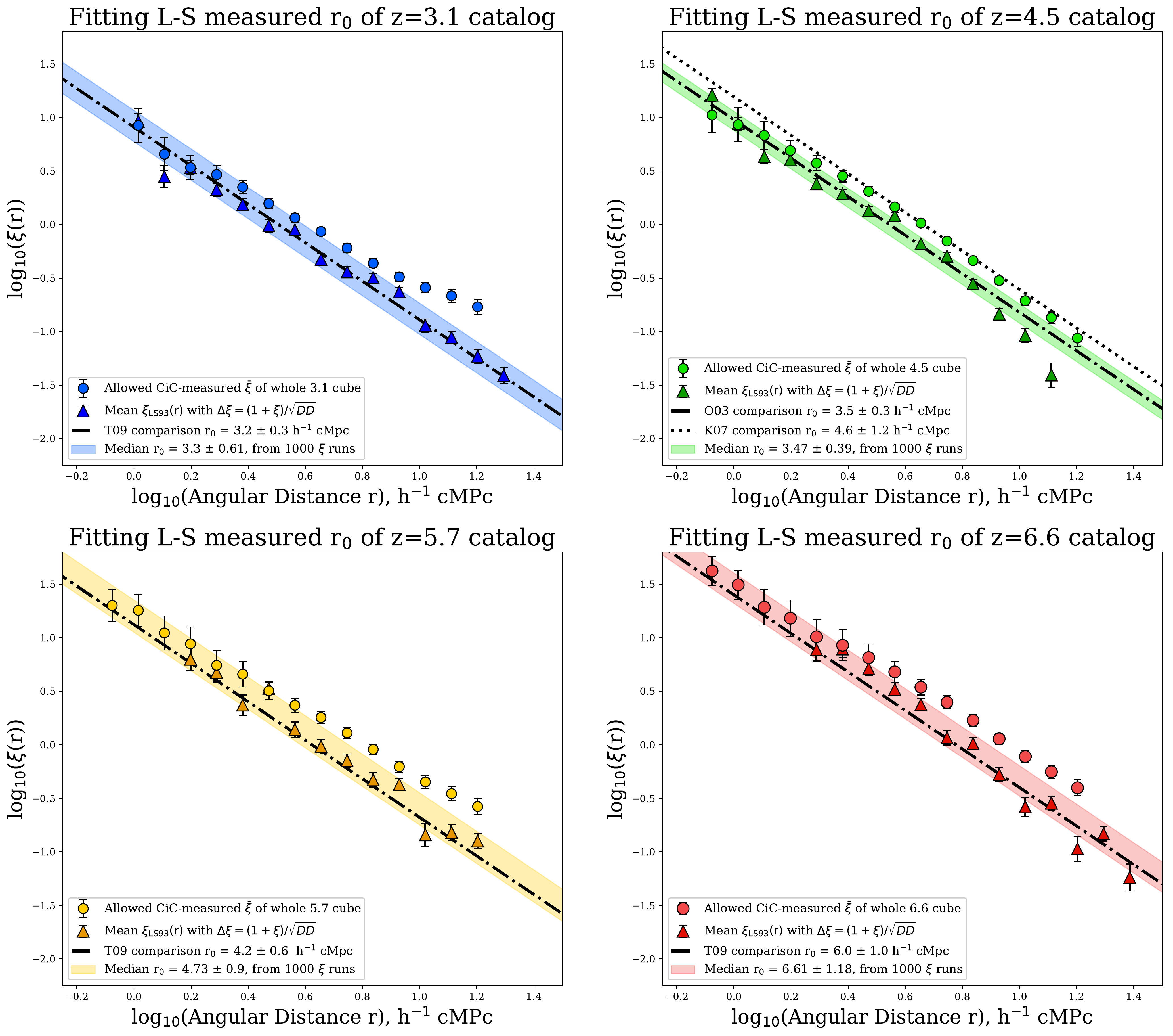}
	\caption{The spatial 3D two-point correlation function of the LAE catalogs and their best power law fits. We calculate the standard two point correlation function with the \citet{L-S1993} estimator 1000 times with new random catalogs, and measure the correlation length of each iteration. We use a least-square method that minimizes the difference between the measured $\xi_{\text{L-S}}(r)$ and the one predicted from the power law in Equation \ref{eq:xi_slope}, assuming a fixed slope of 1+$\delta$=1.8. The darker colored triangles indicate the mean $\xi_{\text{L-S}}$ value across the 1000 iterations at each radius. The error bars on the colored triangles are the 1$\sigma$ logarithmic space Poisson error, $\Delta \xi = (1+\xi)/\sqrt{DD}$, the irreducible error from the size of the sample. 
	\\
	The black dash-dot line marks the best-fit correlation length and error from \citet{Tilvi2009} for the $z=3.1,\ 5.7,\ 6.6$ catalogs, which agree within 1$\sigma$ errors with our new correlation length measurements. The $z=4.5$ LAE catalog had an output error undiscovered in T09 that led to excess clustering signals, so we ignore the affected regions when measuring the clustering. We find great agreement within $1\sigma$ errors with the correlation length of observed LAEs at $z=4.5$ from K07 and O03. The final $1\sigma$ errors in our correlation lengths come from adding in quadrature the median fitting error across the 1000 calculated correlation lengths and one standard deviation in the correlation length distribution (Figure \ref{fig:HistLinFitsRo}). The colored shaded regions and legend entries are the $\pm 3\sigma$ region about the median $r_0\ h^{-1}$ cMpc. The lighter colored circles are the count-in-cells measured volume-averaged two-point correlation function $\bar{\xi}_{\text{CiC}}(r)$, which are used for the VPF transformations later in this work. The higher amplitude is expected with volume-averaged correlation functions, and the pattern of higher clustering at higher redshifts remains. }

	\label{fig:TilviLinFitsRo}
	\end{center}
\end{figure*}

\begin{figure*}
	\begin{center}
    \includegraphics[width=.9\textwidth]{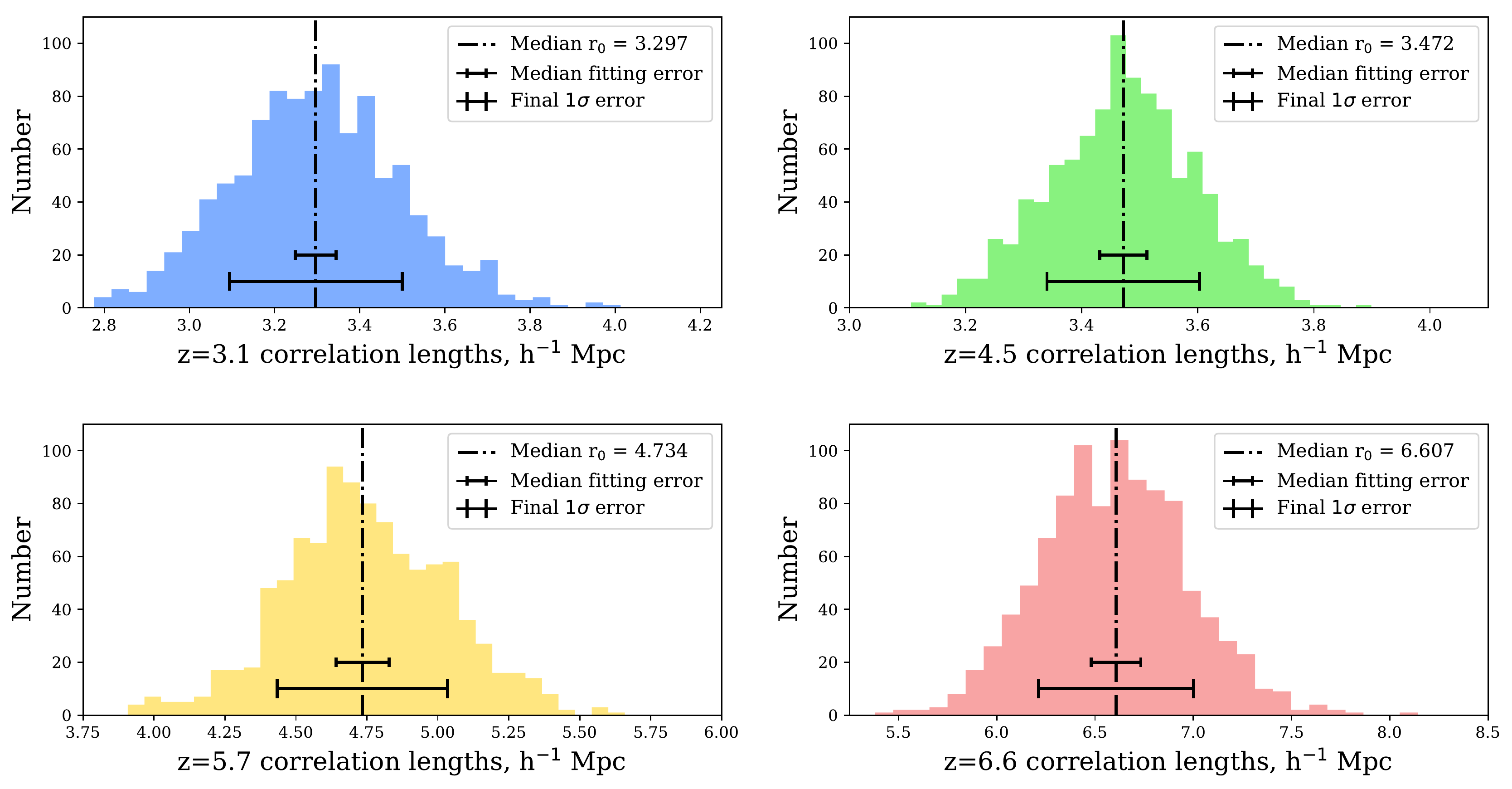}
	\caption{The correlation length fits to the spatial 3D two-point correlation function of the LAE catalogs. We calculate the two point correlation function $\xi$ with the \citet{L-S1993} estimator 1000 times with new random catalogs, and measure the correlation length of each iteration. The median fitting error from our least squares fitting method is added in quadrature to the $1\sigma$ standard deviation across each $r_0$ distribution to measure a $1\sigma$ error on the median correlation length value. The $z=3.1,\ 5.7,\ 6.6$ catalogs agree very well with the previous measurement from T09 and their comparison correlation lengths.}

	\label{fig:HistLinFitsRo}
	\end{center}
\end{figure*}

The two-point correlation function measures the excess probability of finding a galaxy at a given separation $r$ from another, relative to a random Poisson distribution (\citealt{PeeblesLSStextbook, CoilReview2013}):

\begin{equation}
    dP = n [ 1 + \xi(r) ] dV,
\end{equation} 

where $n$ is the mean number density of the sample. The two-point correlation function is measured in three dimensions in comoving space, and yields the power spectrum with a Fourier Transform (see \citealt{PeeblesLSStextbook} for a thorough text on large scale structure). We use the three dimensional form of the two-point correlation function estimator from \citet{L-S1993}, where the data and randomly distributed point catalogs are the same size:

\begin{equation}
\xi_{\text{L-S}}(r) = \frac{DD(r) - 2 DR(r) + RR(r)}{RR(r)}
\end{equation}

The two-point correlation function of galaxy samples is known to mostly follow a power law with a slope of 1+$\delta$, with the long-used value of $\delta$=0.77 $\pm$ .006 from \citet{Peebles1975} and \citet{Totsuji1969}. The amplitude of the power law gives the correlation length $r_0$, which is defined as the scale length of clustering for a given galaxy sample when $\xi=1$. The correlation length is often expressed in terms of $h^{-1}$ cMpc to facilitate comparison between observations with different cosmological assumptions. A larger correlation length value roughly indicates more clustering. To calculate the best fit correlation length for each catalog, we fit this power law on the $\log_{10}(r)$ (h$^{-1}$ cMpc) vs. $\log_{10}(\xi_{\text{L-S}}(r))$ plot for the two-point correlation function: 

\begin{equation}
\xi = \Big( \frac{r_0}{r} \Big) ^{1+\delta} \approx \Big( \frac{r_0}{r} \Big) ^{1.8}
\label{eq:xi_slope}
\end{equation}

T09 previously calculated the two-point correlation functions and measured correlation lengths for their simulated LAEs, but to maintain transparency and account for small differences in luminosity cuts in our analysis, we repeat the measurements here. We verify the accuracy of the calculation and correlation length fit by finding great agreement with T09's correlation lengths for the $z=3.1, 5.7, 6.6$ catalogs. Due to the previously undetected output error in the $z=4.5$ catalog in T09, we ignore the affected corner and measure the clustering the rest of the volume. We compare directly to the observed LAEs of K07 and O03 and find much better agreement with their correlation length measurements.

First, we calculate each catalog's `data-data' distance counts ($DD(r)$): for each data point, we measure three-dimensional distance $d = \sqrt{x^2 + y^2 + z^2}$ to every other data point and organized each distance into the corresponding radius bin. Then, for the $DR$ and $RR$ terms, we generate random points in the same way as the VPF random circles were (but here, matching the number of galaxies). For each random point, we compare distance to other random points (creating the $RR(r)$ array) and to the data points (creating the $DR(r)$ array). We repeat the $DR$ and $RR$ tallying process with 1000 different random point sets, as to eventually have 1000 unique $\xi_{\text{L-S}}(r)$ measurements for each catalog. 

In Figure \ref{fig:TilviLinFitsRo}, we plot the median $\xi_{\text{L-S}}$ across all 1000 repetitions with traditional Poisson errors: $\Delta \xi = (1+\xi)/ \sqrt{DD}$. This is an irreducible error in the correlation function. As \citet{Khostovan2018a} found for their emission line galaxy samples, the Poisson error is consistent with the computationally expensive bootstrapping technique for galaxy samples between $10^2$ and $10^3$. Additionally, we plot the volume-averaged two-point correlation function $\bar{\xi}$ we measure in \textsection \ref{subsec:CiCMethod} using count-in-cells in colored circles, along with the Poisson error. The transition to the volume-averaged two-point correlation function from the traditional form (eg. \citealt{L-S1993}) increases the amplitude of the power law by approximately $\log_{10}(3/(2-\beta)$ (regardless of the volume scale, if assuming an ideal power law of slope $1+\beta$).\footnote{For those curious about moving from $\bar{\xi}_2$ to the standard $\xi_2$, see the derivation in \citet{YangSaslaw2011}.} Because the two-point correlation function is not an ideal power law and drops off at the smallest and largest scales, the $\bar{\xi}$ measurement from CiC will approximate a power law with a less drastic amplitude increase. CiC was shown to be consistent with the volume average of the \citet{L-S1993} estimator by \citet{Szapudi1998}, and we confirm this in our LAEs.

To best approximate the value and error on the correlation lengths for the samples, we fit each of the 1000 individual $\xi_{\text{L-S}}$ curves to a power law\footnote{We use \textit{scipy.optimize.leastsq} \citep{Scipy}} with a fixed slope of 1+$\delta$=1.8 and choose the median value of the resulting distribution (Figure \ref{fig:HistLinFitsRo}). We add the distribution's $1\sigma$ standard deviation in quadrature to the median error out of the least squares fitting algorithm covariance matrix to get our final correlation length $1\sigma$ error. In Figure \ref{fig:TilviLinFitsRo}, the color shaded regions and the values listed in the legends correspond to the median $r_0 \pm 3\sigma\ h^{-1}$ cMpc.
The newest correlation lengths, whose 3$\sigma$ range is shown as shaded regions and described in the legend of Figure \ref{fig:TilviLinFitsRo}, agree very well with the T09 calculations and/or those of comparison observations. The new and previous correlation length measurements for $z=3.1, 5.7, 6.6$ agree within all 1$\sigma$ errors. For the $z=4.5$ catalog, the previously undetected empty back corner created a discrepancy between the T09 correlation length and those which K07 and O03 measured in observed LAEs. The new correlation length measurements for $z=4.5$ agree within all 1$\sigma$ errors with the O03 and K07 measurements.

\subsection{Hierarchical Scaling and Higher-order Clustering Moments}
\label{subsec:HierarScaling}

Different galaxy distributions can have identical two-point correlation functions but unique VPFs that include additional information from higher order correlation functions (\citealt{Maurogordato1987, White1979}). In the framework of hierarchical scaling, one assumes that all the volume averaged correlation functions are hierarchically related to the volume-averaged two-point correlation function ($\bar{\xi}_2$) via the hierarchical \textit{Ansatz}:

\begin{equation}
\bar{\xi}_p = S_p\ \bar{\xi}_2^{\ p-1}, \quad p \geq 3;  \qquad \text{where} \quad \bar{\xi}_p (R) = \frac{ \int \xi_p(r) dV_R}{ \int dV_R} .
\label{eq:hierarchy}
\end{equation}

This \textit{Ansatz} allows one to transform the VPF of \citet{White1979}, now expressing it as a sum of all the higher order volume-averaged correlation functions, their scaling coefficients, and the average number of galaxies in a tested volume of radius R, $\bar{N}(r)$:

\begin{equation}
P_0(r) = \exp \Bigg( \sum^{\infty}_{p=1} \frac{- \bar{N}(r)^p}{p!}\  \bar{\xi}_p(r) \Bigg) \quad \rightarrow \quad
P_0(r) = \exp \Bigg( \sum^{\infty}_{p=1} \frac{- \bar{N}(r)^p}{p!}\ S_p\  \bar{\xi}_{2}^{p-1}(r) \Bigg).
\label{eq:VPFhierarchical}
\end{equation}

One can derive the $S_p$ scaling coefficients by assuming a phenomenological model of hierarchical gravitational clustering (Poisson, Gaussian, negative binomial, thermodynamic, etc.). Then, one is able to connect all the correlation functions to the VPF with just the $S_p$ coefficients. This applies only when both are three dimensional and while in redshift space (\citealt{Kaiser1987, Ryden1996}). These phenomenological models predict specific strengths for all the correlation functions, and deviations from these predictions when compared to measured void statistics or higher order correlation functions are used to explore the underlying physics of clustering (eg. \citealt{Fry1989}).

The hierarchical scaling models can connect the VPF to the volume-averaged two-point correlation function and higher-order correlation functions. The \textit{reduced void probability function} scales the VPF by the average number of galaxies in cells of a given size:

\begin{equation}
\chi(r) = \frac{-\ln(P_0(r))}{\bar{N}(r)}.
\label{eq:def_rVPF}
\end{equation} 

Measurements of $\bar{\xi}$, $\chi$, and $\bar{N}$ can be used to determine which hierarchical scaling model best describes how the clustering of galaxies occurs. Different hierarchical scaling models provide different solutions to the cosmological many-body problem with varying physical justifications, and determining which best describe galaxy clustering can probe the fundamental physics behind the formation and evolution of large scale structure (eg. \citealt{SaslawFang1996}). In Appendix A we summarize all the models against which we test our simulated LAE catalogs. We focus on the negative binomial model when further testing how the VPF transforms back and forth from count-in-cells.

\subsection{The Negative Binomial Model for Hierarchical Scaling}
\label{subsec:NegBinModelDef}

The negative binomial model (NBM) for the hierarchical equations that govern gravitational clustering has been found to be the best fitting model for $z<1$ galaxy samples with complete spectroscopic redshift coverage (C05; Cr04; \citealt{Maurogordato1987}; \citealt{Gaztanaga+Yokoyama1993}; \citealt{Tinker2008}; \citealt{Bel2016}; \citealt{YangSaslaw2011}; \citealt{Hurtado-Gil2017}; among several others) and in simulations (\citealt{Andrew2013}). The negative binomial model predicts a reduced VPF ($\chi$) from the volume-averaged two-point correlation function ($\bar{\xi}_2$) and the average number of galaxies in a cell ($\bar{N}$):

\begin{equation}
\chi_{\text{neg. bin.}} = \frac{\ln(1+\bar{N}\bar{\xi}_2)}{\bar{N}\bar{\xi}_2}
\label{eq:NegBin_Xi}
\end{equation}

The NBM can be derived through many methods and has been used to statistically describe phenomena across many fields of science. It is also known as the modified Bose-Einstein distribution (see \citealt{Fry2013} for a succinct derivation in that context). The NBM was first used a cosmological context by \citet{C-D1983} and derived for clustering analysis by \citet{E-G1992}. In this reworking, the probability of a galaxy appearing in a given cell depends on the number of galaxies that already exist within it, and is correlated to a uniform Poisson distribution. Or as Cr04 summarizes it, the NBM describes the probability of having a given number of `successes' (finding a galaxy) after a certain number of `failures' (voids), and probability of a failure ($P_0$) depends on the density of a sample and how clustered it is ($\bar{N}\bar{\xi}_2$). The NBM was also derived with thermodynamic arguments by \citet{Sheth1995} in the framework of \citet{SaslawHamilton1984}, and shown to be a special case of the hyper geometric model of \citet{Mekjian2007}. Most recently, \citet{Gaztanaga+Yokoyama1993} rederived the model by considering a sample divided into equal and independent cells and tying the cells' occupation probability to $\bar{\xi}$.

Although it is an robust description of many galaxy samples, the NBM is arguably not physically motivated. \citet{Fry2013} argue that ``there is no fundamental reason that galaxies follow the negative binomial scaling curve, but that this follows from typical galaxy parameters'' like bias and number density. Some authors find it justified (\citealt{C-D1983}; \citealt{E-G1992}; \citealt{Betancort-Rijo2000}). \citet{SaslawFang1996} argued the NBM violates the second law of thermodynamics while still being the best fit to their data. \citet{YangSaslaw2011} confirmed that their SDSS sample was consistent with both the NBM and quasi-equilibrium model, and preferred the quasi-equilibrium model for its physical explanation. Additionally, \citet{YangSaslaw2011} found that the large cosmic variance and underestimated error from jackknife errors could explain why the negative binomial function best fit a very similar sample in C05. Later, \citet{Hurtado-Gil2017} confirmed that the NBM outperformed the quasi-equilibrium and other models in a blind fit of the SDSS galaxies after careful consideration of incompleteness and noise in the sample.

\begin{figure*}
  \centering
\includegraphics[width=\textwidth]{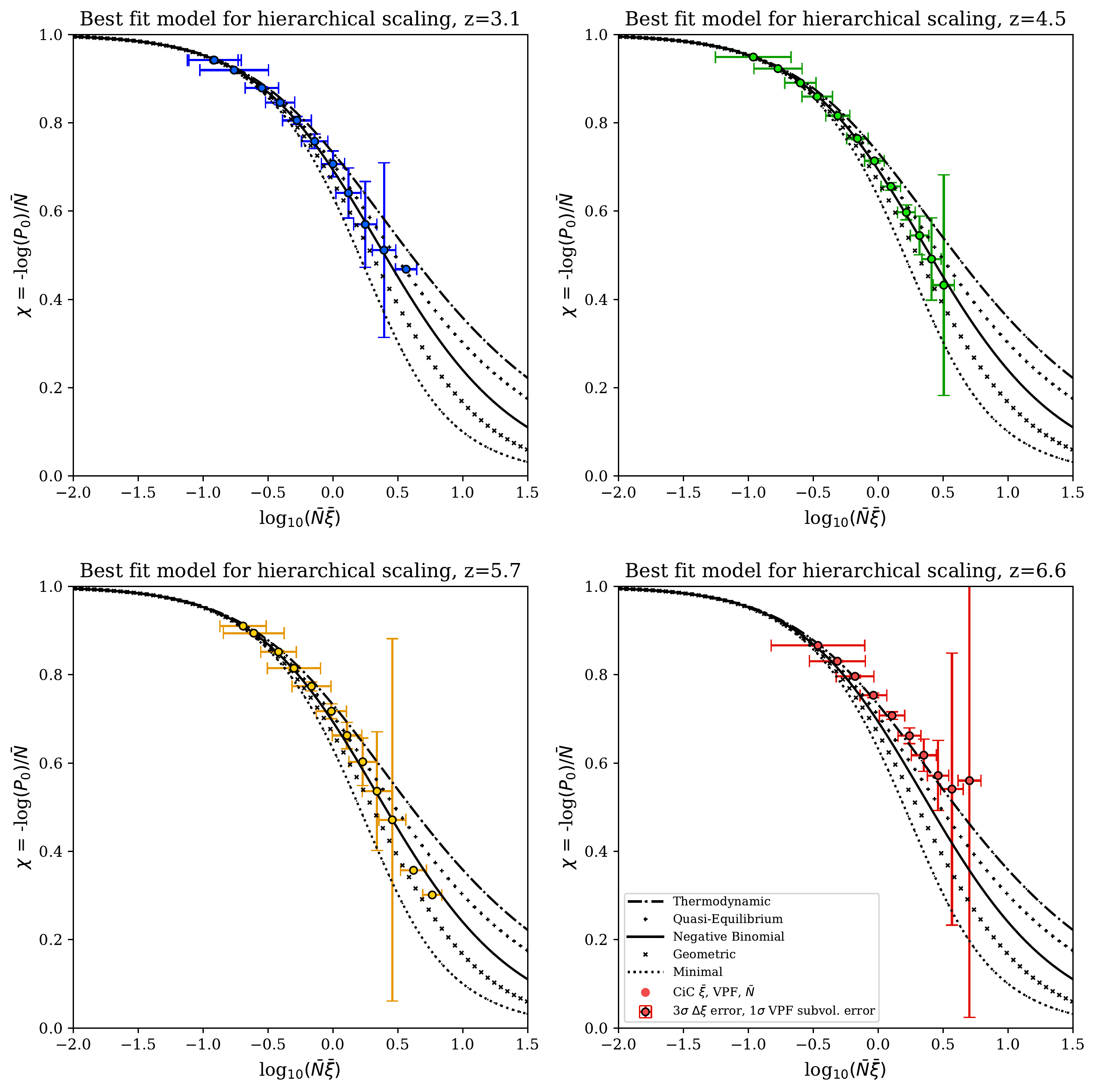}
     \caption{The reduced VPF, $\chi=-\log(P_0)/\bar{N}$, as a function of $\bar{N}\bar{\xi}$ for several hierarchical scaling models and our catalogs of simulated LAEs. The negative binomial model is the best fit for the $z=3.1, 4.5, 5.7$ catalogs, as expected, though the quasi-equilibrium model is not ruled out. The $z=6.6$ catalog is perhaps most consistent with the thermodynamic model, but the large $1\sigma$ errors from the VPF subsample analysis for this smaller sample do not discount the negative binomial model or quasi-equilibrium model as the best fit. For errors, we assume that $\bar{N}$ has no uncertainty, and we transfer the errors on the 3D VPF and $\xi_{\text{L-S}}$ to $\log(\chi)$ and $\log(\bar{N}\bar{\xi})$ respectively. The $y$-errors are the $1\sigma$ VPF subsample error (the shaded regions in the right side of Figure 1), and the $x$-errors are 3 times the $\Delta \xi = (1+\xi_{\text{L-S}})/\sqrt{DD}$ errors (the error bars in Figure 2) scaled to $\log_{10}(\bar{\xi})$. For a completely unclustered Poisson sample, $\chi$ = 1 everywhere. The consistency of the z=3.1, 4.5, and 5.7 catalogs with the negative binomial model supports the hypothesis that LAEs also follow the negative binomial model. The deviation of the LAEs at z=6.6 from the NBM and its significance requires larger simulations for confirmation at a significant .}
    \label{fig:HierarModels}
\end{figure*}

\section{Fitting and Testing the NBM for LAEs} 
\label{sec:FitTestNBM}

\subsection{Fitting Count-In-Cells to the NBM}
\label{subsec:CiCMethod}
  
Traditionally, the reduced VPF ($\chi$) and volume-averaged correlation functions ($\bar{\xi}_p$) are measured using the count-in-cells method. 
For CiC, one drops many random test spheres of a given size and counts the number of galaxies within each. The VPF can be measured with CiC by isolating the empty test spheres. The volume-averaged two-point correlation function, $\bar{\xi}_2$, is the reduced second moment of the probability distribution of the galaxies' CiC, or the variance in the number of galaxies across all test spheres $i$ of a given size:

\begin{equation}
    \bar{\xi}_2 = \frac{ \overline{ (N_i - \bar{N})^2 }  - \bar{N}}{\bar{N}^2} 
\label{eq:CiC_barXi}
\end{equation}

As with $\bar{\xi}_2$, the higher order terms $\bar{\xi}_p$ are the reduced $p$'th order moments, and CiC is often used to measure the scaling coefficients $S_p$ between the correlation functions (\citealt{Croton2004b}; \citealt{Wolk2013}; among many others). CiC has also been used to help constrain cosmological models in simulations and observations (\citealt{Wang2019, Uhlemann2020, Wen2020, ReppSzapudi2020}.) CiC is a versatile probe for the underlying causes of the hierarchical features that we see in large scale structure, and is also often used in other broader applications of clustering (eg. \citealt{Adelberger1998, Mesinger+Furlanetto2008, Jensen2014}). As we continue this analysis, we consider how we can leverage what we have learned about the VPF's uncertainty and reliability when using CiC.

To calculate our CiC, we drop at least 500,000 test spheres at every radius, utilizing the same random placement as we used for the simple VPF and the \citet{L-S1993} $\xi$ measurements. We confirm that the CiC algorithm gives identical VPF values to our previous measurement when counting empty cells, and therefore should give accurate $\bar{N}$ and $\bar{\xi}$ measurements as well. Additionally, we calculate the value of $\bar{\xi}$ at the largest radii we measured and found this integral constraint negligibly small. 

The more traditional application of hierarchical scaling models has been to use CiC to plot $\bar{N}\bar{\xi}$ against the reduced VPF $\chi$ and determine what model of hierarchical scaling best fits the galaxy samples, as in C05 and Cr04. In Figure \ref{fig:HierarModels}, we plot $\log_{10}(\bar{N}\bar{\xi})$ vs. $\chi$ for our catalogs and the predicted relationships from several popular models, compiled in Appendix A from \citealt{Fry2013}, C05, and Cr04. We assume zero uncertainty in $\bar{N}$, translate the $1\sigma$ subsample-method VPF errors to $\chi(r)$, and translate the $3\sigma\ \Delta \xi_{\text{L-S}}$ errors to $\log_{10}(\bar{N}\bar{\xi})$. By transferring our subsample method error estimation for the VPF into the reduced VPF, we can better inform the uncertainty of CiC, which might be underestimated by jackknife sampling \citep{YangSaslaw2011}.

In agreement with the work in Cr04, C05, \citealt{Andrew2013}, \citealt{Hurtado-Gil2017}, and many others, we find the NBM is the best fit model of hierarchical scaling for the $z=3.1, 4.5, 5.7$ simulated LAEs. These catalogs' $1\sigma$ errors on $\chi$, though, would not discount the geometric model or the gravitational quasi-equilibrium model (which is preferred over the NBM for its physical motivation in \citealt{YangSaslaw2011}). The $z=6.6$ catalog might be more consistent with the thermodynamic model, which treats galaxy clustering by analogy to statistical mechanics (\citealt{SaslawHamilton1984}; \citealt{Fry1986}), but its large $1\sigma$ VPF errors do not discount the negative binomial or gravitational quasi-equilibrium models.

\subsection{Testing our VPF with the NBM and $\bar{\xi}_{\text{CiC}}$} 
\label{subsec:Xibar_to_VPF}

\begin{figure*}
  \centering
    \includegraphics[width=.88\textwidth]{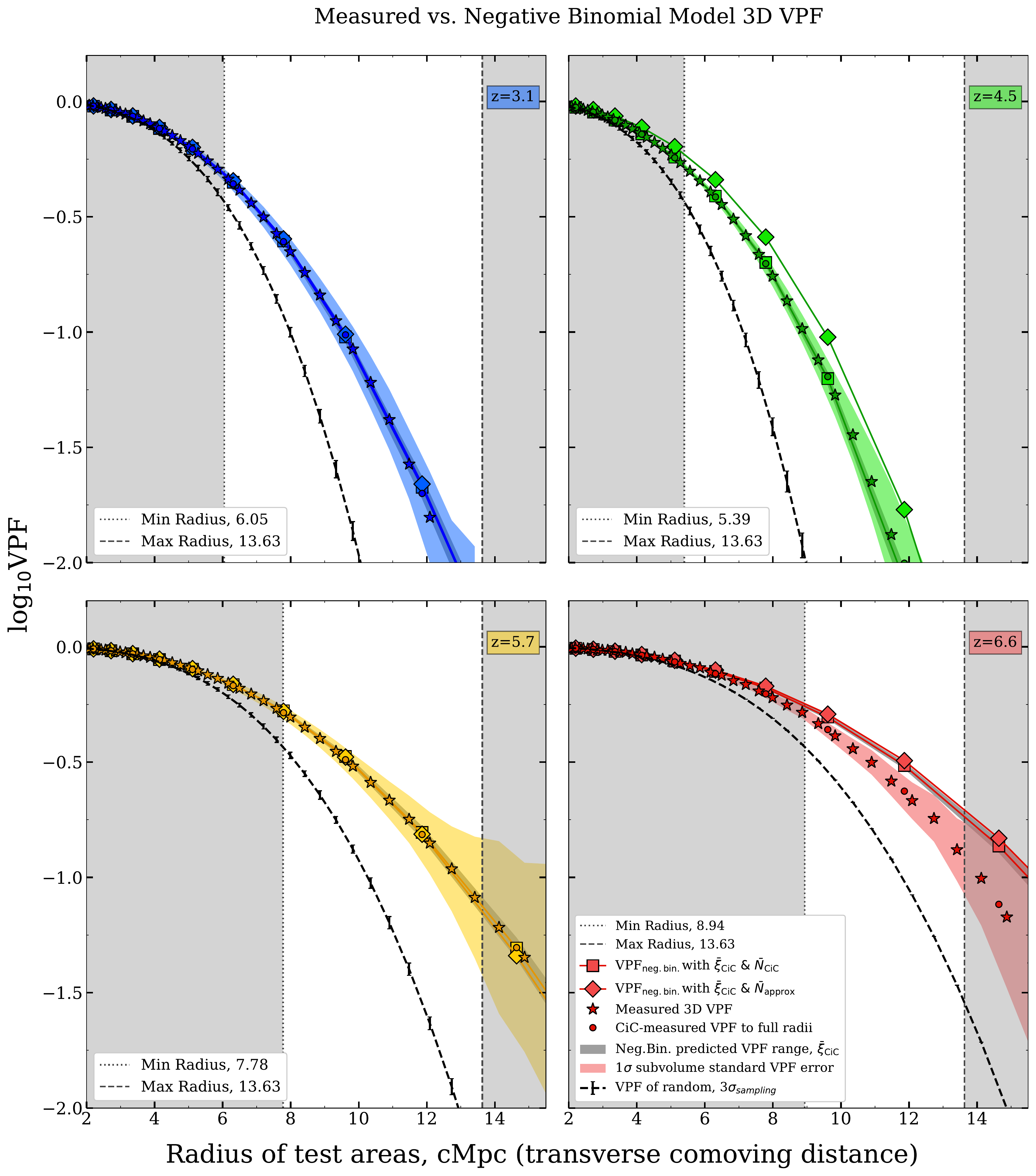}
    \caption{Comparing the measured 3D VPF vs. the VPF derived from the negative binomial model using the measured count-in-cells volume averaged two-point correlation function. The dark colored stars are the measured 3D VPF of each volume, and the shaded colored regions indicate the $1\sigma$ standard error across the independent (51 cMpc)$^3$ sub-volumes' VPFs. The dark colored circles are the VPF measured with the CiC, and the excellent match with our independent measurement confirms our algorithm is correct. The colored squares (diamonds) are the derived VPF using $\bar{\xi}_{\text{CiC}}$ and $\bar{N}_{\text{CiC}}$ ($\bar{N}_{\text{approx}}$). Our final results do not change with our choice for $z=3.1, 5.7, 6.6$, and very slightly change for $z=4.5$, indicating that approximating $\bar{N}$ from the number density is is a valid shortcut to measuring it with CiC. The grey shading around the $\bar{\xi}_{\text{CiC}}\ \&\ \bar{N}_{\text{CiC}}$ squares comes from carrying $\pm 1\sigma = \Delta\xi=(1+\xi)/\sqrt{DD}$ errors from $\xi_{\text{L-S}}$ through the negative binomial model. The black dashed lines indicate the predicted VPF for the simulations were they completely randomly distributed. The maximum and minimum radii are theoretically derived from the simulations' number densities, volumes, and our chosen VPF precision of $\pm 10^{-2}$. As Figures \ref{fig:HierarModels} and \ref{fig:ComparingNegBinXi} also show, the negative binomial model transformations using the CiC agree excellently with the observed clustering signals for $z=3.1, 4.5, 5.7$. The z=6.6 LAEs deviate from the negative binomial model predictions, lying just past the $1\sigma$ error bounds of the measurements.}
    \label{fig:ComparingNegBinVPF}

\end{figure*}

To further test how well the negative binomial model fits our simulated LAEs, we use it to move between the VPF and volume-averaged two-point correlation function of our simulated LAEs. We first start with the CiC-measured $\bar{\xi}$, use the NBM to predict a VPF, and compare to the measured VPF as a test of the model fit. We combine the definitions of the reduced VPF and the NBM in Equations \ref{eq:def_rVPF} and \ref{eq:NegBin_Xi} to yield the following form of the model, which we can solve with various root-seeking algorithms at each radius.\footnote{We use \textit{scipy.optimize.root} and the Levenberg-Marquardt sub-module \citep{Scipy} for its stability, and verify the answers with manual evaluations. We also remove abnormally large values generated at numerically unstable regions, and limit ourselves to $|\bar{\xi}_{\text{VPF}}| < 3$.} This transformation allows us to derive a volume averaged two-point correlation function from the VPF of the T09 simulated LAEs, or vice versa:

\begin{equation}
0 = P_0\ (1 + \bar{N}\bar{\xi}_2)^{1/\bar{\xi}_2} -1
\label{eq:solveforXibar}
\end{equation}

In their work, C05 measured the VPF, $\bar{\xi}$, and $\bar{N}$ with CiC. To further separate the VPF from the CiC and two-point correlation function, we approximate $\bar{N}(r)$ (and therefore $\chi$) by assuming homogeneous distribution at the same number density of the catalogs:

\begin{equation}
\bar{N} _{\text{approx}} (r) \approx \mathcal{N}\ \frac{4}{3} \pi R^3 \quad \rightarrow \quad
\chi(r) \approx \frac{-\ln(P_0(r))}{\mathcal{N}\ (4/3) \pi R^3}.
\label{eq:Nbardef}
\end{equation}

We note that our shortcut to the reduced VPF is reminiscent of the normalized VPF value explored by \citet{Maurogordato1987}, in which $n$ is the number volume density and $V$ is any given volume:

\begin{equation}
    \chi_{\text{M-L,1987}}= \frac{\ln(P_0)}{nV}
\end{equation}

We find that this $\bar{N}_{\text{approx}}$ is significantly similar to the true measured value of $\bar{N}_{\text{CiC}}$, and our final results are not affected by this choice. This similarity is to be expected, as both attempt to measure an average of galaxies in a given volume, though $\bar{N}_{\text{approx}}$ and $\bar{N}_{\text{CiC}}$ might diverge in situations where the shape of the one-point count-in-cells distribution is very skewed (as in some of the cosmological Quijote simulations tested in \citealt{Uhlemann2020} using the probability density function of the entire matter field).  For us, the only notable difference between $\bar{N}_{\text{approx}}$ and $\bar{N}_{\text{CiC}}$ is in the $z=4.5 $ catalog, where using $\bar{N}_{\text{approx}}$ raises log$_{10}(\bar{\xi}_{\text{VPF}})$ by about the width of the 1$\sigma$ VPF errors in Figure \ref{fig:ComparingNegBinVPF}.

First, we follow the typical application of the hierarchical scaling models to derive a VPF from the CiC measurements and then compare to the measured VPF in Figure \ref{fig:ComparingNegBinVPF}. We assume a NBM to transform the volume averaged two-point correlation function into a VPF, the more traditional use of hierarchical scaling as seen in C05 and Cr04. We compare how well the volume averaged form from CiC ($\bar{\xi}_{\text{CiC}}$) recreates the VPF under the NBM, with either the true measured average number of galaxies from CiC ($\bar{N}_{\text{CiC}}$) or an approximation from number density ($\bar{N}_{\text{approx}}$). As Figure \ref{fig:HierarModels} predicted, using all CiC measurements for $\bar{\xi}$ and $\bar{N}$ reproduces the VPF exactly for the $z=3.1, 4.5, 5.7$ catalogs. We verify that the shortcut of using $\bar{N}_{\text{approx}}$ over $\bar{N}_{\text{CiC}}$ does not influence the final conclusions. The deviation of the CiC values from the NBM in $z=6.6$ lie at the edge large 1$\sigma$ errors about the VPF measurement, nearly overlapping with the transformed 1$\sigma\ \Delta \xi_{\text{L-S}}$ errors about the NBM-predicted VPF. 

\subsection{Deriving $\bar{\xi}$ from the VPF with NBM}

Past studies like those of C05, Cr04, and \citet{Andrew2013} measured two point and higher order correlation functions, derived the VPF assuming given models for the hierarchical scaling, and then compared against measured VPF to compare the models. Our work here builds off their key results--the strength of the NBM to predict the VPF of low-redshift galaxies--and inverts this pattern in our simulations of high-redshift LAEs as an additional test of the NBM. We begin from a measured VPF, assume the NBM to derive a correlation function, and then compare to the measured CiC correlation function. This serves as an extra step to test the NBM, the behavior of CiC for different samples, and the shortcut of using a density-approximated $\bar{N}_{\text{approx}}$ over the directly CiC-measured $\bar{N}_{\text{CiC}}$. 
We find that deviations from NBM, especially in the $z=6.6$ catalog, appear more obvious when moving from the VPF into $\bar{\xi}$ than in the opposite direction, thanks to the transformation into straight lines in log-log space.

\begin{figure*}
  \centering
    \includegraphics[width=\textwidth]{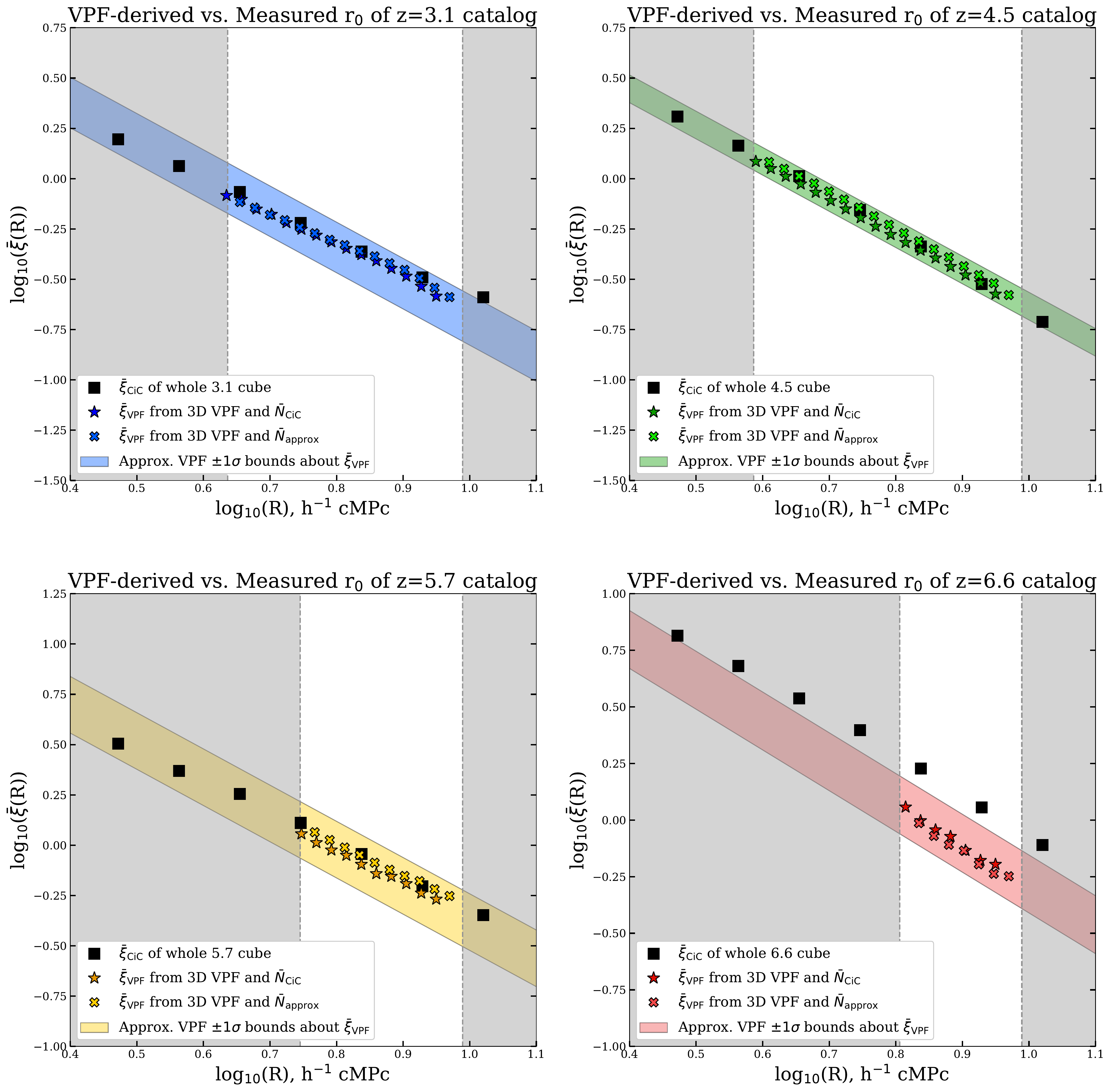}
    \caption{
    Comparing the directly measured $\bar{\xi}_{\text{CiC}}$ vs. the $\bar{\xi}_{\text{VPF}}$ from transforming the VPF using the negative binomial model.
    The black squares are $\bar{\xi}_{\text{CiC}}$ calculated using Equation \ref{eq:CiC_barXi}. The colored stars (crosses) are the derived $\bar{\xi}_{\text{VPF}}$ from the central 3D VPF of the simulations, using the CiC-measured (density-approximated) $\bar{N}$. The colored shaded regions correspond the power law fits to the $\bar{\xi}_{\text{VPF}}$ from the VPF $\pm 1\sigma$ subsample uncertainties, and straddle the best-fit power law for our central $\bar{\xi}_{\text{VPF}}$ values.
    We calculate $\bar{\xi}_{\text{VPF}}$ within the theoretical radii derived and discussed in \textsection \ref{subsec:TheoryErroronVPF} (grey regions), based on the catalogs' number densities, volumes, and our chosen precision of $\pm 10^{-2}$. The pattern of clustering between the catalogs translates through $\bar{\xi}_{\text{VPF}}$.
    \\
    The $\bar{\xi}_{\text{VPF}}$ matches the $\bar{\xi}_{\text{CiC}}$ very well for the $z=3.1,4.5,5.7$ catalogs, confirming the negative binomial is the best fitting scaling model for those catalogs. The $z=6.6$ catalog $\bar{\xi}_{\text{VPF}}$ and $\bar{\xi}_{\text{CiC}}$ appear to disagree here (perhaps due to this catalog's size and strong clustering), though Figure \ref{fig:HierarModels} and the $1\sigma$ VPF bounds suggest this deviation from the negative binomial model might not be statistically significant. The choice of $\bar{N}$ makes little difference to measuring $\bar{\xi}$ with $\bar{\xi}_{\text{VPF}}$, meaning that the number density approximation with the VPF can serve as a shortcut to measuring CiC for large samples.
    }
    \label{fig:ComparingNegBinXi}
\end{figure*}

We solve the NBM in Equation \ref{eq:solveforXibar} for $\bar{\xi}_{\text{VPF}}$ in Figure \ref{fig:ComparingNegBinXi} given our measured VPF $P_0$ and either $\bar{N}_{\text{CiC}}$ or $\bar{N}_{\text{approx}}$. We compare the derived $\bar{\xi}_{\text{VPF}}$ (colored stars and crosses, depending on the choice of $\bar{N}$) and the CiC-measured $\bar{\xi}_{\text{CiC}}$. We show the ranges corresponding to the power law fits to the $\bar{\xi}_{\text{VPF}}$ from the VPF $\pm 1\sigma$ subsample uncertainties, which straddle the best-fit power law for our central $\bar{\xi}_{\text{VPF}}$ values. Figure \ref{fig:ComparingNegBinXi} agrees with Figures \ref{fig:HierarModels} and \ref{fig:ComparingNegBinVPF}: the $z=3.1,4.5,5.7$ catalogs show excellent consistency with the NBM, as their $\bar{\xi}_{\text{VPF}}$ are exactly consistent with $\bar{\xi}_{\text{CiC}}$. The unique behavior of the $z=6.6$ catalog might be clarified here, showing the discrepancy from the NBM in Figure \ref{fig:HierarModels} centers around the LAEs' $\bar{\xi}_{\text{CiC}}$ not behaving as the NBM predicts. This might be due to the fact that the $z=6.6$ catalog is the smallest and most clustered, perhaps indicating that higher clustering amplitudes might make our approximation for $\bar{N}$ become invalid earlier and create larger deviations from the true hierarchical scaling description. The reasons for the different behavior of the z=6.6 LAE catalog require larger samples and larger simulations to confirm and understand. We leave that for future work.

We again confirm that three of the four catalogs show great agreement with the NBM and the last good agreement with the quasi-equilibrium model (the next best-fitting hierarchical scaling model), agreeing with many low-redshift galaxy observations. Therefore, we show that the approximating the average number of galaxies in a cell $\bar{N}$ using number density and assuming the NBM allows us to accurately recreate the CiC volume-averaged two-point correlation function only using the VPF. Additionally, our $\bar{\xi}_{\text{VPF}}$ carry through the intuitive errors that we derived in \textsection \ref{subsec:TheoryErroronVPF}, and avoid the possibly underestimated errors often given to CiC measurements \citep{YangSaslaw2011}. 

\section{Conclusion} \label{sec:Conclusion}

The Void Probability Function is an underutilized measure of clustering that inherently involves the higher order correlation functions (\citealt{White1979, Maurogordato1987}). It contains additional information about large scale structure and higher order clustering that the two-point correlation function cannot discern. When it has been used, the VPF has been able to fit low-redshift galaxy observations to a negative binomial model for the hierarchical gravitational clustering moments (eg. \citealt{Croton2004b, Conroy2005}). For many years, the VPF was treated as a uniquely capable measurement of clustering for very small galaxy samples (\citealt{Palunas2004, Kashikawa2006}), and as a powerful probe for dark matter halo occupation modeling (\citealt{Fry1986, Maurogordato1987, Little&Weinberg1994, Berl&Weinb2002}). In this work, we attempt to determine and test the limits of what the VPF is capable, utilize it to probe higher-order clustering in LAEs, and offer ways it can inform clustering analyses with count-in-cells and the two-point correlation function.

We measure the 2D and 3D void probability function of four simulated catalogs of Lyman-$\alpha$ emitters at redshifts 3.1, 4.5, 5.7, and 6.6 from the work of \citet{Tilvi2009}. We offer general guidelines for understanding and quantifying the uncertainty in the VPF. We present suggested limits to when the VPF can be effectively studied, depending on the number density of the sample, the volume of the survey, and the minimum VPF value to be studied. We choose the independent subsample method to approximate the error in the VPF after finding that jackknife sampling understates the uncertainty of the VPF. We divide each volume into the eight independent cubes of (51 cMpc)$^{3}$ and measure the VPF on each sub-volume and the standard error across them as a $1 \sigma$ error on the VPF. These simulated LAE catalogs and our VPF measurements can serve as external `correct' error estimators for observations of LAEs.

Next, we measure the standard \citet{L-S1993} and volume averaged Count-in-Cells two-point correlation functions for the catalogs. We re-affirm the $z=3.1,5.7,6.6$ and improve the $z=4.5$ correlation lengths measured in T09, and explore the transformation between $\xi$ and $\bar{\xi}$ for our catalogs. Then, we compare how different models of hierarchical scaling between the volume-averaged correlation functions compare to the catalogs' measured $\bar{N}_{\text{CiC}}$, $\bar{\xi}_{\text{CiC}}$, and VPF. We find the $z=3.1,4.5,5.7$ catalogs are best fit by the negative binomial model, though the uncertainties on the VPF do not rule out the quasi-equilibrium model. We find that the $z=6.6$ might be better fit by the thermodynamic or quasi-equilibrium models, though the uncertainties on our VPF values do not discount the negative binomial model.

We further explore how the negative binomial model fits our simulated LAEs by transforming $\bar{\xi}_{\text{CiC}}$ into a VPF, and then the VPF into a $\bar{\xi}_{\text{VPF}}$. We confirm that approximating $\bar{N}$ with the number density rather than using the measured $\bar{N}_{\text{CiC}}$ does not change our final results. We find all four catalogs' $\bar{\xi}_{\text{CiC}}$ predict accurate VPF values with the negative binomial model (within both the $1\sigma$ VPF and $1\sigma\ \Delta\xi$ errors). We find that the $z=3.1,4.5,5.7$ catalogs' VPFs derive very accurate $\bar{\xi}_{\text{VPF}}$ with the NBM, while the $z=6.6$ VPF yields a lower-than-expected $\bar{\xi}_{\text{VPF}}$ (though still within our uncertainties). The deviation of the NBM at our highest redshift might perhaps explained by the fact that $z=6.6$ is our sparsest and most clustered catalog. This behavior requires larger simulations to fully understand. These results indicate that high-redshift starburst galaxies like LAEs in the T09 model show large-scale clustering behavior similar to that of local galaxies, also following the negative binomial model. This suggests the LAEs likely do not have excessive clustering in higher-order correlation terms, 
though the size of our simulations do not let us decisively rule out the presence of higher-order clustering nor identify additional redshift dependence.

Our guidelines of when the VPF is best used offer greater use beyond improving future clustering analyses in moderately sized samples. In this age of growing astronomical data, there are many problems that would respond positively to fast and intuitive tools with well-explored uncertainties. We have shown, as expected, that the VPF conserves the same core qualitative conclusion of the standard \citet{L-S1993} two-point correlation function through the NBM transformation: the clustering of LAEs increases with redshift. 
This overlap allows us to consider which clustering applications might prefer the brevity of calculating the VPF and its inherent connection to higher-order clustering. For example, the upcoming Nancy Grace Roman Space Telescope (previously known as WFIRST) will observe many thousands of LAEs during the Epoch of Reionization ($z>6$). Due to the resonant properties of the \lya\ line in neutral hydrogen, the prevalence and clustering of LAEs is able to track the amount of neutral hydrogen and its distribution around them (eg. \citealt{McQuinn2007}; \citealt{Jensen2013}; \citealt{Zheng2016}; \citealt{Kakiichi2016}; \citealt{Ouchi2018}). Samples in more neutral IGM will be more clustered, both because their emission is more attenuated at higher neutral fractions, but also because LAEs are more likely to be observed in regions that have already been ionized (eg. \citealt{Furlanetto2006}). For this application, the VPF might be a useful tool (\citealt{Gangolli2020}, Perez et al. in prep)--it easily compares clustering between samples of the same density, is more efficient to calculate than CiC, and has intuitive limits that can help guide the planning of observations.

\acknowledgments
\textit{We thank Evan Scannapieco, Rob Thacker, Ilian Iliev, and Garrelt Mellema for their contributions to the \citet{Tilvi2009} simulations that underpin our calculations. We also deeply thank the anonymous referee for their comments and feedback, which greatly improved this manuscript.
The first year of this research was supported by a PhD Science Fellowship from the National GEM Consortium through Arizona State University, and afterwards supported the School of Earth $\&$ Space Exploration at Arizona State University and NSF grant AST-1518057.  This work was carried out on the \textit{Saguaro} supercomputing cluster operated by the Fulton School of Engineering at Arizona State University, and also made use of NASA's Astrophysics Data System Bibliographic Services. Our figures were made with \textit{Matplotlib} \citep{Matplotlib}, and the bulk of our calculations used \textit{Numpy} \citep{Numpy}, \textit{Scipy} \citep{Scipy}, and the \citet{NedWright} cosmology calculator. }

\appendix
We summarize the various models of hierarchical scaling between the volume-averaged correlation functions that we test for our simulated LAE catalogs. We discuss the negative binomial model in great detail in \textsection \ref{subsec:NegBinModelDef}. The bulk of this list and the functional forms of the models come from C05, Cr04, and \citet{Fry2013}. The order of these models follows the order of their curves in Figure \ref{fig:HierarModels}, from top to bottom.
\\

\subsection{Thermodynamic and Gravitational Quasi-Equilibrium Models}

(The naming schemes for these models have evolved over time, and what we call the QEM is referred to as the thermodynamic model in several places. We identify the QEM as \citealt{Fry2013} do, and call the final form of the model that \citealt{Fry1986} derives the thermodynamic model.)

The thermodynamic (black dash-dot curve) and the so-called gravitational quasi-equilibrium (black crosses) in \ref{fig:HierarModels} are both derived from the same occupation probability distribution of  \citet{SaslawHamilton1984}. They developed a thermodynamic theory of the properties of gravitational clustering that yields this occupation distribution \citep{Fry1986}:

\begin{equation}
    P_N = \frac{\bar{N}(1-b)}{N!} [\bar{n}(1-b) + Nb]^{N-1} \exp (-\bar{N}(1-b) - Nb ).
\end{equation}

A thorough explanation of the derivation of the two models can be found in \citet{Fry1986}, which we summarize here. The ``quasi-equilibrium'' model (black crosses in Figure \ref{fig:HierarModels}) comes from approximation discrete realization of a continuous background number density:

\begin{equation}
    \chi_{\mathrm{QEM}} = (1+\bar{N}\bar{\xi})^{-1/2}
\end{equation}

One can then apply a continuum limit to process the $P_N$ distribution into a discretized continuum for large $\bar{N}$:

\begin{equation}
    N^{N-2}Q_N = (2N-3)!!, \quad K=\bar{\xi}^{-1}[1-(1-2\bar{N}\bar{\xi})^{1/2}],
\end{equation}

which gives the form of the ``thermodynamic'' model we plot as a black dash-dot curve in \ref{fig:HierarModels}:

\begin{equation}
    \chi_{\mathrm{thermo}} = \Big( (1+2\bar{N}\bar{\xi})^{1/2} -1 \Big) / \bar{N}\bar{\xi}
\end{equation}

\citet{Ahmad2002} derived the thermodynamic model from statistical mechanics, imagining that galaxy clustering evolves through a sequence of quasi-equilibrium states and supplying fundamental justification for the model. \citet{Sheth1998} derived a model that yields the same $S_p$ scaling coefficients as the thermodynamic model, but by instead approaching the evolution of the dark matter halo mass function as an extension of the excusion set model. \citet{Fry1985} details that the thermodynamic model is only valid in large volumes (once $\bar{N}\bar{\xi}$ has converged), and expresses concern at how such large scales could have become thermodynamically relaxed over age of universe. (Given that at the redshifts of our simulated LAEs, the universe was 800 million to 2.1 billion years old, this concern remains.)

\subsection{Geometric Model}

This model begins with an occupancy probability distribution $p_n \propto p^n$, and is a specialized case of the hyper-geometric model of \citet{Mekjian2007}, the \citet{Hamilton1988} model, and alternate form of the model in \citet{Alimi1990}. The geometric hierarchical model takes the form:

\begin{equation}
    \chi_{\mathrm{geom}} = 1/(1 + \frac{1}{2}\bar{N}\bar{\xi})    
\end{equation}

\subsection{Minimal Model}

The minimal model can be derived by taking the moments of galaxy counts in a hierarchical Poisson model, in which galaxies form in randomly places clusters of $N_c$ galaxies and $\bar{N}\bar{\xi} \gg 1$ \citep{Fry1988}. This model is another limiting case of the \citet{Mekjian2007} model. The minimal Poisson model has all scaling coefficients $S_p=1$ for all $p$, and takes the form:

\begin{equation}
    \chi_{\mathrm{min}} = \Big( 1 - \exp(-\bar{N}\bar{\xi})  \Big) / \bar{N}\bar{\xi}
\end{equation}

\bibliography{references}

\begin{thebibliography}{}
\expandafter\ifx\csname natexlab\endcsname\relax\def\natexlab#1{#1}\fi
\providecommand{\url}[1]{\href{#1}{#1}}

\bibitem[{{Adelberger} {et~al.}(1998){Adelberger}, {Steidel}, {Giavalisco},
  {Dickinson}, {Pettini}, \& {Kellogg}}]{Adelberger1998}
{Adelberger}, K.~L., {Steidel}, C.~C., {Giavalisco}, M., {et~al.} 1998, \apj,
  505, 18

\bibitem[{{Ahmad} {et~al.}(2002){Ahmad}, {Saslaw}, \& {Bhat}}]{Ahmad2002}
{Ahmad}, F., {Saslaw}, W.~C., \& {Bhat}, N.~I. 2002, \apj, 571, 576

\bibitem[{{Alimi} {et~al.}(1990){Alimi}, {Blanchard}, \&
  {Schaeffer}}]{Alimi1990}
{Alimi}, J.-M., {Blanchard}, A., \& {Schaeffer}, R. 1990, \apjl, 349, L5

\bibitem[{{Andrew} {et~al.}(2013){Andrew}, {Barnaby}, \& {Taylor}}]{Andrew2013}
{Andrew}, K., {Barnaby}, D., \& {Taylor}, L. 2013, ArXiv e-prints,
  arXiv:1312.6648

\bibitem[{{Atek} {et~al.}(2014){Atek}, {Kunth}, {Schaerer}, {Mas-Hesse},
  {Hayes}, {{\"O}stlin}, \& {Kneib}}]{Atek2014}
{Atek}, H., {Kunth}, D., {Schaerer}, D., {et~al.} 2014, \aap, 561, A89

\bibitem[{{Baugh} {et~al.}(2004){Baugh}, {Croton}, {Gazta{\~n}aga}, {Norberg},
  {Colless}, {Baldry}, {Bland -Hawthorn}, {Bridges}, {Cannon}, {Cole},
  {Collins}, {Couch}, {Dalton}, {De Propris}, {Driver}, {Efstathiou}, {Ellis},
  {Frenk}, {Glazebrook}, {Jackson}, {Lahav}, {Lewis}, {Lumsden}, {Maddox},
  {Madgwick}, {Peacock}, {Peterson}, {Sutherland}, {Taylor}, \& {2dFGRS
  Team}}]{Baugh2004}
{Baugh}, C.~M., {Croton}, D.~J., {Gazta{\~n}aga}, E., {et~al.} 2004, \mnras,
  351, L44

\bibitem[{{Bel} {et~al.}(2016){Bel}, {Branchini}, {Di Porto}, {Cucciati},
  {Granett}, {Iovino}, {de la Torre}, {Marinoni}, {Guzzo}, {Moscardini},
  {Cappi}, {Abbas}, {Adami}, {Arnouts}, {Bolzonella}, {Bottini}, {Coupon},
  {Davidzon}, {De Lucia}, {Fritz}, {Franzetti}, {Fumana}, {Garilli}, {Ilbert},
  {Krywult}, {Le Brun}, {Le F{\`e}vre}, {Maccagni}, {Ma{\l}ek}, {Marulli},
  {McCracken}, {Paioro}, {Polletta}, {Pollo}, {Schlagenhaufer}, {Scodeggio},
  {Tasca}, {Tojeiro}, {Vergani}, {Zanichelli}, {Burden}, {Marchetti},
  {Mellier}, {Nichol}, {Peacock}, {Percival}, {Phleps}, \& {Wolk}}]{Bel2016}
{Bel}, J., {Branchini}, E., {Di Porto}, C., {et~al.} 2016, \aap, 588, A51

\bibitem[{{Beltz-Mohrmann} {et~al.}(2020){Beltz-Mohrmann}, {Berlind}, \&
  {Szewciw}}]{BeltzMohrmann2020}
{Beltz-Mohrmann}, G.~D., {Berlind}, A.~A., \& {Szewciw}, A.~O. 2020, \mnras,
  491, 5771

\bibitem[{{Benson}(2010)}]{Benson2010}
{Benson}, A.~J. 2010, \physrep, 495, 33

\bibitem[{{Berlind} \& {Weinberg}(2002)}]{Berl&Weinb2002}
{Berlind}, A.~A., \& {Weinberg}, D.~H. 2002, \apj, 575, 587

\bibitem[{{Bernardeau} {et~al.}(2002){Bernardeau}, {Colombi}, {Gazta{\~n}aga},
  \& {Scoccimarro}}]{Bernardeau2002}
{Bernardeau}, F., {Colombi}, S., {Gazta{\~n}aga}, E., \& {Scoccimarro}, R.
  2002, \physrep, 367, 1

\bibitem[{Betancort-Rijo(2000)}]{Betancort-Rijo2000}
Betancort-Rijo, J. 2000, Journal of Statistical Physics, 98, 917.
\newblock \url{https://doi.org/10.1023/A:1018691912596}

\bibitem[{{Bielby} {et~al.}(2016){Bielby}, {Tummuangpak}, {Shanks}, {Francke},
  {Crighton}, {Ba{\~n}ados}, {Gonz{\'a}lez-L{\'o}pez}, {Infante}, \&
  {Orsi}}]{Bielby2016}
{Bielby}, R.~M., {Tummuangpak}, P., {Shanks}, T., {et~al.} 2016, \mnras, 456,
  4061

\bibitem[{{Carruthers} \& {Duong-van}(1983)}]{C-D1983}
{Carruthers}, P., \& {Duong-van}, M. 1983, Physics Letters B, 131, 116

\bibitem[{{Coil}(2013)}]{CoilReview2013}
{Coil}, A.~L. 2013, {The Large-Scale Structure of the Universe}, Vol.~6
  (Springer Science+Business Media Dordrecht), 384

\bibitem[{{Conroy} {et~al.}(2005){Conroy}, {Coil}, {White}, {Newman}, {Yan},
  {Cooper}, {Gerke}, {Davis}, \& {Koo}}]{Conroy2005}
{Conroy}, C., {Coil}, A.~L., {White}, M., {et~al.} 2005, \apj, 635, 990

\bibitem[{{Cooper} {et~al.}(2007){Cooper}, {Newman}, {Coil}, {Croton}, {Gerke},
  {Yan}, {Davis}, {Faber}, {Guhathakurta}, {Koo}, {Weiner}, \&
  {Willmer}}]{Cooper2007}
{Cooper}, M.~C., {Newman}, J.~A., {Coil}, A.~L., {et~al.} 2007, \mnras, 376,
  1445

\bibitem[{{Crocce} {et~al.}(2006){Crocce}, {Pueblas}, \&
  {Scoccimarro}}]{LagrangePerturb_Crocce}
{Crocce}, M., {Pueblas}, S., \& {Scoccimarro}, R. 2006, \mnras, 373, 369

\bibitem[{{Croton} {et~al.}(2004{\natexlab{a}}){Croton}, {Colless},
  {Gazta{\~n}aga}, {Baugh}, {Norberg}, {Baldry}, {Bland-Hawthorn}, {Bridges},
  {Cannon}, {Cole}, {Collins}, {Couch}, {Dalton}, {de Propris}, {Driver},
  {Efstathiou}, {Ellis}, {Frenk}, {Glazebrook}, {Jackson}, {Lahav}, {Lewis},
  {Lumsden}, {Maddox}, {Madgwick}, {Peacock}, {Peterson}, {Sutherland}, \&
  {Taylor}}]{Croton2004b}
{Croton}, D.~J., {Colless}, M., {Gazta{\~n}aga}, E., {et~al.}
  2004{\natexlab{a}}, \mnras, 352, 828

\bibitem[{{Croton} {et~al.}(2004{\natexlab{b}}){Croton}, {Gazta{\~n}aga},
  {Baugh}, {Norberg}, {Colless}, {Baldry}, {Bland -Hawthorn}, {Bridges},
  {Cannon}, {Cole}, {Collins}, {Couch}, {Dalton}, {De Propris}, {Driver},
  {Efstathiou}, {Ellis}, {Frenk}, {Glazebrook}, {Jackson}, {Lahav}, {Lewis},
  {Lumsden}, {Maddox}, {Madgwick}, {Peacock}, {Peterson}, {Sutherland}, \&
  {Taylor}}]{Croton2004a}
{Croton}, D.~J., {Gazta{\~n}aga}, E., {Baugh}, C.~M., {et~al.}
  2004{\natexlab{b}}, \mnras, 352, 1232

\bibitem[{{Croton} {et~al.}(2005){Croton}, {Farrar}, {Norberg}, {Colless},
  {Peacock}, {Baldry}, {Baugh}, {Bland-Hawthorn}, {Bridges}, {Cannon}, {Cole},
  {Collins}, {Couch}, {Dalton}, {De Propris}, {Driver}, {Efstathiou}, {Ellis},
  {Frenk}, {Glazebrook}, {Jackson}, {Lahav}, {Lewis}, {Lumsden}, {Maddox},
  {Madgwick}, {Peterson}, {Sutherland}, \& {Taylor}}]{Croton2005}
{Croton}, D.~J., {Farrar}, G.~R., {Norberg}, P., {et~al.} 2005, \mnras, 356,
  1155

\bibitem[{{Davis} {et~al.}(1985){Davis}, {Efstathiou}, {Frenk}, \&
  {White}}]{FoFhalofinder}
{Davis}, M., {Efstathiou}, G., {Frenk}, C.~S., \& {White}, S.~D.~M. 1985, \apj,
  292, 371

\bibitem[{{Dayal} {et~al.}(2010){Dayal}, {Ferrara}, \& {Saro}}]{Dayal2010}
{Dayal}, P., {Ferrara}, A., \& {Saro}, A. 2010, \mnras, 402, 1449

\bibitem[{{Dayal} {et~al.}(2009){Dayal}, {Ferrara}, {Saro}, {Salvaterra},
  {Borgani}, \& {Tornatore}}]{Dayal2009}
{Dayal}, P., {Ferrara}, A., {Saro}, A., {et~al.} 2009, \mnras, 400, 2000

\bibitem[{{Dayal} {et~al.}(2011){Dayal}, {Maselli}, \& {Ferrara}}]{Dayal2011}
{Dayal}, P., {Maselli}, A., \& {Ferrara}, A. 2011, \mnras, 410, 830

\bibitem[{{Elizalde} \& {Gaztanaga}(1992)}]{E-G1992}
{Elizalde}, E., \& {Gaztanaga}, E. 1992, \mnras, 254, 247

\bibitem[{{Finkelstein} {et~al.}(2008){Finkelstein}, {Rhoads}, {Malhotra},
  {Grogin}, \& {Wang}}]{Finkelstein2008}
{Finkelstein}, S.~L., {Rhoads}, J.~E., {Malhotra}, S., {Grogin}, N., \& {Wang},
  J. 2008, \apj, 678, 655

\bibitem[{{Finkelstein} {et~al.}(2007){Finkelstein}, {Rhoads}, {Malhotra},
  {Pirzkal}, \& {Wang}}]{Finkelstein2007}
{Finkelstein}, S.~L., {Rhoads}, J.~E., {Malhotra}, S., {Pirzkal}, N., \&
  {Wang}, J. 2007, \apj, 660, 1023

\bibitem[{{Fry}(1985)}]{Fry1985}
{Fry}, J.~N. 1985, \apj, 289, 10

\bibitem[{{Fry}(1986)}]{Fry1986}
---. 1986, \apj, 306, 358

\bibitem[{{Fry}(1988)}]{Fry1988}
---. 1988, Publications of the Astronomical Society of the Pacific, 100, 1336

\bibitem[{{Fry} \& {Colombi}(2013)}]{Fry2013}
{Fry}, J.~N., \& {Colombi}, S. 2013, \mnras, 433, 581

\bibitem[{{Fry} {et~al.}(1989){Fry}, {Giovanelli}, {Haynes}, {Melott}, \&
  {Scherrer}}]{Fry1989}
{Fry}, J.~N., {Giovanelli}, R., {Haynes}, M.~P., {Melott}, A.~L., \&
  {Scherrer}, R.~J. 1989, \apj, 340, 11

\bibitem[{{Furlanetto} {et~al.}(2006){Furlanetto}, {Zaldarriaga}, \&
  {Hernquist}}]{Furlanetto2006}
{Furlanetto}, S.~R., {Zaldarriaga}, M., \& {Hernquist}, L. 2006, \mnras, 365,
  1012

\bibitem[{{Gangolli} {et~al.}(submitted){Gangolli}, {D'Aloisio}, {Nasir}, \&
  {Zheng}}]{Gangolli2020}
{Gangolli}, N., {D'Aloisio}, A., {Nasir}, F., \& {Zheng}, Z. submitted, \mnras

\bibitem[{{Gawiser} {et~al.}(2007){Gawiser}, {Francke}, {Lai}, {Schawinski},
  {Gronwall}, {Ciardullo}, {Quadri}, {Orsi}, {Barrientos}, {Blanc}, {Fazio},
  {Feldmeier}, {Huang}, {Infante}, {Lira}, {Padilla}, {Taylor}, {Treister},
  {Urry}, {van Dokkum}, \& {Virani}}]{Gawiser2007}
{Gawiser}, E., {Francke}, H., {Lai}, K., {et~al.} 2007, \apj, 671, 278

\bibitem[{{Gaztanaga} \& {Yokoyama}(1993)}]{Gaztanaga+Yokoyama1993}
{Gaztanaga}, E., \& {Yokoyama}, J. 1993, \apj, 403, 450

\bibitem[{{Gronwall} {et~al.}(2007){Gronwall}, {Ciardullo}, {Hickey},
  {Gawiser}, {Feldmeier}, {van Dokkum}, {Urry}, {Herrera}, {Lehmer}, {Infante},
  {Orsi}, {Marchesini}, {Blanc}, {Francke}, {Lira}, \&
  {Treister}}]{Gronwall2007}
{Gronwall}, C., {Ciardullo}, R., {Hickey}, T., {et~al.} 2007, \apj, 667, 79

\bibitem[{{Guaita} {et~al.}(2010){Guaita}, {Gawiser}, {Padilla}, {Francke},
  {Bond}, {Gronwall}, {Ciardullo}, {Feldmeier}, {Sinawa}, {Blanc}, \&
  {Virani}}]{Guaita2010}
{Guaita}, L., {Gawiser}, E., {Padilla}, N., {et~al.} 2010, \apj, 714, 255

\bibitem[{{Hamilton}(1988)}]{Hamilton1988}
{Hamilton}, A.~J.~S. 1988, \apj, 332, 67

\bibitem[{{Hao} {et~al.}(2018){Hao}, {Huang}, {Xia}, {Zheng}, {Jiang}, \&
  {Li}}]{Hao2018}
{Hao}, C.-N., {Huang}, J.-S., {Xia}, X., {et~al.} 2018, \apj, 864, 145

\bibitem[{{Harikane} {et~al.}(2018){Harikane}, {Ouchi}, {Shibuya}, {Kojima},
  {Zhang}, {Itoh}, {Ono}, {Higuchi}, {Inoue}, {Chevallard}, {Capak}, {Nagao},
  {Onodera}, {Faisst}, {Martin}, {Rauch}, {Bruzual}, {Charlot}, {Davidzon},
  {Fujimoto}, {Hilmi}, {Ilbert}, {Lee}, {Matsuoka}, {Silverman}, \&
  {Toft}}]{Harikane2018}
{Harikane}, Y., {Ouchi}, M., {Shibuya}, T., {et~al.} 2018, \apj, 859, 84

\bibitem[{{Hong} {et~al.}(2019){Hong}, {Dey}, {Lee}, {Orsi}, {Gebhardt},
  {Vogelsberger}, {Hernquist}, {Xue}, {Jung}, {Finklestein}, {Tuttle}, \&
  {Boylan- Kolchin}}]{Hong2019}
{Hong}, S., {Dey}, A., {Lee}, K.-S., {et~al.} 2019, \mnras, 483, 3950

\bibitem[{{Hunter}(2007)}]{Matplotlib}
{Hunter}, J. 2007, {{Matplotlib}: A 2D Graphics Environment}, 90--95

\bibitem[{{Hurtado-Gil} {et~al.}(2017){Hurtado-Gil}, {Mart{\'\i}nez}, {Arnalte-
  Mur}, {Pons-Border{\'\i}a}, {Pareja-Flores}, \& {Paredes}}]{Hurtado-Gil2017}
{Hurtado-Gil}, L., {Mart{\'\i}nez}, V.~J., {Arnalte- Mur}, P., {et~al.} 2017,
  \aap, 601, A40

\bibitem[{{Iliev} {et~al.}(2008){Iliev}, {Shapiro}, {McDonald}, {Mellema}, \&
  {Pen}}]{Iliev2008}
{Iliev}, I.~T., {Shapiro}, P.~R., {McDonald}, P., {Mellema}, G., \& {Pen},
  U.-L. 2008, \mnras, 391, 63

\bibitem[{{Inoue} {et~al.}(2018){Inoue}, {Hasegawa}, {Ishiyama}, {Yajima},
  {Shimizu}, {Umemura}, {Konno}, {Harikane}, {Shibuya}, {Ouchi}, {Shimasaku},
  {Ono}, {Kusakabe}, {Higuchi}, \& {Lee}}]{Inoue2018}
{Inoue}, A.~K., {Hasegawa}, K., {Ishiyama}, T., {et~al.} 2018, \pasj, 70, 55

\bibitem[{{Jensen} {et~al.}(2014){Jensen}, {Hayes}, {Iliev}, {Laursen},
  {Mellema}, \& {Zackrisson}}]{Jensen2014}
{Jensen}, H., {Hayes}, M., {Iliev}, I.~T., {et~al.} 2014, \mnras, 444, 2114

\bibitem[{{Jensen} {et~al.}(2013){Jensen}, {Laursen}, {Mellema}, {Iliev},
  {Sommer-Larsen}, \& {Shapiro}}]{Jensen2013}
{Jensen}, H., {Laursen}, P., {Mellema}, G., {et~al.} 2013, \mnras, 428, 1366

\bibitem[{Jones {et~al.}(2001--)Jones, Oliphant, Peterson, {et~al.}}]{Scipy}
Jones, E., Oliphant, T., Peterson, P., {et~al.} 2001--, {SciPy}: Open source
  scientific tools for {Python},  Scipy.
\newblock \url{http://www.scipy.org/}

\bibitem[{{Kaiser}(1987)}]{Kaiser1987}
{Kaiser}, N. 1987, \mnras, 227, 1

\bibitem[{{Kakiichi} {et~al.}(2016){Kakiichi}, {Dijkstra}, {Ciardi}, \&
  {Graziani}}]{Kakiichi2016}
{Kakiichi}, K., {Dijkstra}, M., {Ciardi}, B., \& {Graziani}, L. 2016, \mnras,
  463, 4019

\bibitem[{{Kashikawa} {et~al.}(2006){Kashikawa}, {Shimasaku}, {Malkan}, {Doi},
  {Matsuda}, {Ouchi}, {Taniguchi}, {Ly}, {Nagao}, {Iye}, {Motohara},
  {Murayama}, {Murozono}, {Nariai}, {Ohta}, {Okamura}, {Sasaki}, {Shioya}, \&
  {Umemura}}]{Kashikawa2006}
{Kashikawa}, N., {Shimasaku}, K., {Malkan}, M.~A., {et~al.} 2006, \apj, 648, 7

\bibitem[{{Khostovan} {et~al.}(2018{\natexlab{a}}){Khostovan}, {Sobral},
  {Mobasher}, {Matthee}, {Cochrane}, {Chartab Soltani}, {Jafariyazani},
  {Paulino-Afonso}, {Santos}, \& {Calhau}}]{Khostovan2018b}
{Khostovan}, A.~A., {Sobral}, D., {Mobasher}, B., {et~al.} 2018{\natexlab{a}},
  arXiv e-prints, arXiv:1811.00556

\bibitem[{{Khostovan} {et~al.}(2018{\natexlab{b}}){Khostovan}, {Sobral},
  {Mobasher}, {Best}, {Smail}, {Matthee}, {Darvish}, {Nayyeri}, {Hemmati}, \&
  {Stott}}]{Khostovan2018a}
---. 2018{\natexlab{b}}, \mnras, 478, 2999

\bibitem[{{Kobayashi} {et~al.}(2007){Kobayashi}, {Totani}, \&
  {Nagashima}}]{Kobayashi2007}
{Kobayashi}, M. A.~R., {Totani}, T., \& {Nagashima}, M. 2007, \apj, 670, 919

\bibitem[{{Kobayashi} {et~al.}(2010){Kobayashi}, {Totani}, \&
  {Nagashima}}]{Kobayashi2010}
---. 2010, \apj, 708, 1119

\bibitem[{{Kova{\v c}} {et~al.}(2007){Kova{\v c}}, {Somerville}, {Rhoads},
  {Malhotra}, \& {Wang}}]{Kovac2007}
{Kova{\v c}}, K., {Somerville}, R.~S., {Rhoads}, J.~E., {Malhotra}, S., \&
  {Wang}, J. 2007, \apj, 668, 15

\bibitem[{{Kusakabe} {et~al.}(2018){Kusakabe}, {Shimasaku}, {Ouchi},
  {Nakajima}, {Goto}, {Hashimoto}, {Konno}, {Harikane}, {Silverman}, \&
  {Capak}}]{Kusakabe2018b}
{Kusakabe}, H., {Shimasaku}, K., {Ouchi}, M., {et~al.} 2018, Publications of
  the Astronomical Society of Japan, 70, 4

\bibitem[{{Landy} \& {Szalay}(1993)}]{L-S1993}
{Landy}, S.~D., \& {Szalay}, A.~S. 1993, \apj, 412, 64

\bibitem[{{Little} \& {Weinberg}(1994)}]{Little&Weinberg1994}
{Little}, B., \& {Weinberg}, D.~H. 1994, \mnras, 267, 605

\bibitem[{{Malhotra} \& {Rhoads}(2002)}]{Malhotra+Rhoads2002}
{Malhotra}, S., \& {Rhoads}, J.~E. 2002, \apj, 565, L71

\bibitem[{{Malhotra} \& {Rhoads}(2004)}]{Malhotra+Rhoads2004}
---. 2004, \apjl, 617, L5

\bibitem[{{Malhotra} {et~al.}(2012){Malhotra}, {Rhoads}, {Finkelstein},
  {Hathi}, {Nilsson}, {McLinden}, \& {Pirzkal}}]{Malhotra2012}
{Malhotra}, S., {Rhoads}, J.~E., {Finkelstein}, S.~L., {et~al.} 2012, \apjl,
  750, L36

\bibitem[{{Matthee} {et~al.}(2016){Matthee}, {Sobral}, {Oteo}, {Best}, {Smail},
  {R{\"o}ttgering}, \& {Paulino-Afonso}}]{Matthee2016}
{Matthee}, J., {Sobral}, D., {Oteo}, I., {et~al.} 2016, \mnras, 458, 449

\bibitem[{{Matthee} {et~al.}(2014){Matthee}, {Sobral}, {Swinbank}, {Smail},
  {Best}, {Kim}, {Franx}, {Milvang-Jensen}, \& {Fynbo}}]{Matthee2014}
{Matthee}, J. J.~A., {Sobral}, D., {Swinbank}, A.~M., {et~al.} 2014, \mnras,
  440, 2375

\bibitem[{{Maurogordato} \& {Lachieze-Rey}(1987)}]{Maurogordato1987}
{Maurogordato}, S., \& {Lachieze-Rey}, M. 1987, \apj, 320, 13

\bibitem[{{McQuinn} {et~al.}(2007){McQuinn}, {Hernquist}, {Zaldarriaga}, \&
  {Dutta}}]{McQuinn2007}
{McQuinn}, M., {Hernquist}, L., {Zaldarriaga}, M., \& {Dutta}, S. 2007, \mnras,
  381, 75

\bibitem[{{Mekjian}(2007)}]{Mekjian2007}
{Mekjian}, A.~Z. 2007, \apj, 655, 1

\bibitem[{{Mesinger} \& {Furlanetto}(2008)}]{Mesinger+Furlanetto2008}
{Mesinger}, A., \& {Furlanetto}, S.~R. 2008, \mnras, 386, 1990

\bibitem[{{Murayama} {et~al.}(2007){Murayama}, {Taniguchi}, {Scoville},
  {Ajiki}, {Sanders}, {Mobasher}, {Aussel}, {Capak}, {Koekemoer}, {Shioya},
  {Nagao}, {Carilli}, {Ellis}, {Garilli}, {Giavalisco}, {Kitzbichler}, {Le
  F{\`e}vre}, {Maccagni}, {Schinnerer}, {Smol{\v{c}}i{\'c}}, {Tribiano},
  {Cimatti}, {Komiyama}, {Miyazaki}, {Sasaki}, {Koda}, \&
  {Karoji}}]{Murayama2007}
{Murayama}, T., {Taniguchi}, Y., {Scoville}, N.~Z., {et~al.} 2007, \apjs, 172,
  523

\bibitem[{{Nagamine} {et~al.}(2010){Nagamine}, {Ouchi}, {Springel}, \&
  {Hernquist}}]{Nagamine2010}
{Nagamine}, K., {Ouchi}, M., {Springel}, V., \& {Hernquist}, L. 2010, \pasj,
  62, 1455

\bibitem[{{Nagashima} \& {Yoshii}(2004)}]{NagashimaYoshii2004}
{Nagashima}, M., \& {Yoshii}, Y. 2004, \apj, 610, 23

\bibitem[{{Nakajima} {et~al.}(2012){Nakajima}, {Ouchi}, {Shimasaku}, {Ono},
  {Lee}, {Foucaud}, {Ly}, {Dale}, {Salim}, {Finn}, {Almaini}, \&
  {Okamura}}]{Nakajima2012}
{Nakajima}, K., {Ouchi}, M., {Shimasaku}, K., {et~al.} 2012, \apj, 745, 12

\bibitem[{{Norberg} {et~al.}(2009){Norberg}, {Baugh}, {Gazta{\~n}aga}, \&
  {Croton}}]{Norberg2009}
{Norberg}, P., {Baugh}, C.~M., {Gazta{\~n}aga}, E., \& {Croton}, D.~J. 2009,
  \mnras, 396, 19

\bibitem[{{Oliphant}(2006)}]{Numpy}
{Oliphant}, T. 2006, {A Guide to {NumPy}}, ed. T.~Publishing

\bibitem[{{Otto} {et~al.}(1986){Otto}, {Politzer}, {Preskill}, \&
  {Wise}}]{Otto1986}
{Otto}, S., {Politzer}, H.~D., {Preskill}, J., \& {Wise}, M.~B. 1986, \apj,
  304, 62

\bibitem[{{Ouchi} {et~al.}(2003){Ouchi}, {Shimasaku}, {Furusawa}, {Miyazaki},
  {Doi}, {Hamabe}, {Hayashino}, {Kimura}, {Kodaira}, {Komiyama}, {Matsuda},
  {Miyazaki}, {Nakata}, {Okamura}, {Sekiguchi}, {Shioya}, {Tamura},
  {Taniguchi}, {Yagi}, \& {Yasuda}}]{Ouchi2003}
{Ouchi}, M., {Shimasaku}, K., {Furusawa}, H., {et~al.} 2003, \apj, 582, 60

\bibitem[{{Ouchi} {et~al.}(2008){Ouchi}, {Shimasaku}, {Akiyama}, {Simpson},
  {Saito}, {Ueda}, {Furusawa}, {Sekiguchi}, {Yamada}, {Kodama}, {Kashikawa},
  {Okamura}, {Iye}, {Takata}, {Yoshida}, \& {Yoshida}}]{Ouchi2008}
{Ouchi}, M., {Shimasaku}, K., {Akiyama}, M., {et~al.} 2008, \apjs, 176, 301

\bibitem[{{Ouchi} {et~al.}(2010){Ouchi}, {Shimasaku}, {Furusawa}, {Saito},
  {Yoshida}, {Akiyama}, {Ono}, {Yamada}, {Ota}, {Kashikawa}, {Iye}, {Kodama},
  {Okamura}, {Simpson}, \& {Yoshida}}]{Ouchi2010}
{Ouchi}, M., {Shimasaku}, K., {Furusawa}, H., {et~al.} 2010, \apj, 723, 869

\bibitem[{{Ouchi} {et~al.}(2018){Ouchi}, {Harikane}, {Shibuya}, {Shimasaku},
  {Taniguchi}, {Konno}, {Kobayashi}, {Kajisawa}, {Nagao}, {Ono}, {Inoue},
  {Umemura}, {Mori}, {Hasegawa}, {Higuchi}, {Komiyama}, {Matsuda}, {Nakajima},
  {Saito}, \& {Wang}}]{Ouchi2018}
{Ouchi}, M., {Harikane}, Y., {Shibuya}, T., {et~al.} 2018, Publications of the
  Astronomical Society of Japan, 70, S13

\bibitem[{{Oyarz{\'u}n} {et~al.}(2017){Oyarz{\'u}n}, {Blanc}, {Gonz{\'a}lez},
  {Mateo}, \& {Bailey}}]{Oyarzun2017}
{Oyarz{\'u}n}, G.~A., {Blanc}, G.~A., {Gonz{\'a}lez}, V., {Mateo}, M., \&
  {Bailey}, John~I., I. 2017, \apj, 843, 133

\bibitem[{{Palunas} {et~al.}(2004){Palunas}, {Teplitz}, {Francis}, {Williger},
  \& {Woodgate}}]{Palunas2004}
{Palunas}, P., {Teplitz}, H.~I., {Francis}, P.~J., {Williger}, G.~M., \&
  {Woodgate}, B.~E. 2004, \apj, 602, 545

\bibitem[{{Partridge} \& {Peebles}(1967)}]{Partridge+Peebles1967}
{Partridge}, R.~B., \& {Peebles}, P.~J.~E. 1967, \apj, 147, 868

\bibitem[{{Peebles}(1975)}]{Peebles1975}
{Peebles}, P.~J.~E. 1975, \apj, 196, 647

\bibitem[{{Peebles}(1980)}]{PeeblesLSStextbook}
---. 1980, {The large-scale structure of the universe}

\bibitem[{{Pirzkal} {et~al.}(2007){Pirzkal}, {Malhotra}, {Rhoads}, \&
  {Xu}}]{Pirzkal2007}
{Pirzkal}, N., {Malhotra}, S., {Rhoads}, J.~E., \& {Xu}, C. 2007, \apj, 667, 49

\bibitem[{{Repp} \& {Szapudi}(2020)}]{ReppSzapudi2020}
{Repp}, A., \& {Szapudi}, I. 2020, arXiv e-prints, arXiv:2006.01146

\bibitem[{{Rhoads} {et~al.}(2000){Rhoads}, {Malhotra}, {Dey}, {Stern},
  {Spinrad}, \& {Jannuzi}}]{Rhoads2000}
{Rhoads}, J.~E., {Malhotra}, S., {Dey}, A., {et~al.} 2000, \apjl, 545, L85

\bibitem[{{Ryden} \& {Melott}(1996)}]{Ryden1996}
{Ryden}, B.~S., \& {Melott}, A.~L. 1996, \apj, 470, 160

\bibitem[{{Santos} {et~al.}(2016){Santos}, {Sobral}, \& {Matthee}}]{Santos2016}
{Santos}, S., {Sobral}, D., \& {Matthee}, J. 2016, \mnras, 463, 1678

\bibitem[{{Saslaw} \& {Fang}(1996)}]{SaslawFang1996}
{Saslaw}, W.~C., \& {Fang}, F. 1996, \apj, 460, 16

\bibitem[{{Saslaw} \& {Hamilton}(1984)}]{SaslawHamilton1984}
{Saslaw}, W.~C., \& {Hamilton}, A.~J.~S. 1984, \apj, 276, 13

\bibitem[{{Sheth}(1995)}]{Sheth1995}
{Sheth}, R.~K. 1995, \mnras, 274, 213

\bibitem[{{Sheth}(1998)}]{Sheth1998}
---. 1998, \mnras, 300, 1057

\bibitem[{{Shimasaku} {et~al.}(2004){Shimasaku}, {Hayashino}, {Matsuda},
  {Ouchi}, {Ohta}, {Okamura}, {Tamura}, {Yamada}, \&
  {Yamauchi}}]{Shimasaku2004}
{Shimasaku}, K., {Hayashino}, T., {Matsuda}, Y., {et~al.} 2004, \apj, 605, L93

\bibitem[{{Shimasaku} {et~al.}(2006){Shimasaku}, {Kashikawa}, {Doi}, {Ly},
  {Malkan}, {Matsuda}, {Ouchi}, {Hayashino}, {Iye}, {Motohara}, {Murayama},
  {Nagao}, {Ohta}, {Okamura}, {Sasaki}, {Shioya}, \&
  {Taniguchi}}]{Shimasaku2006}
{Shimasaku}, K., {Kashikawa}, N., {Doi}, M., {et~al.} 2006, Publications of the
  Astronomical Society of Japan, 58, 313

\bibitem[{{Shioya} {et~al.}(2009){Shioya}, {Taniguchi}, {Sasaki}, {Nagao},
  {Murayama}, {Saito}, {Ideue}, {Nakajima}, {Matsuoka}, {Trump}, {Scoville},
  {Sand ers}, {Mobasher}, {Aussel}, {Capak}, {Kartaltepe}, {Koekemoer},
  {Carilli}, {Ellis}, {Garilli}, {Giavalisco}, {Kitzbichler}, {Impey},
  {LeFevre}, {Schinnerer}, \& {Smolcic}}]{Shioya2009}
{Shioya}, Y., {Taniguchi}, Y., {Sasaki}, S.~S., {et~al.} 2009, \apj, 696, 546

\bibitem[{{Skibba} {et~al.}(2014){Skibba}, {Smith}, {Coil}, {Moustakas},
  {Aird}, {Blanton}, {Bray}, {Cool}, {Eisenstein}, {Mendez}, {Wong}, \&
  {Zhu}}]{Skibba2014}
{Skibba}, R.~A., {Smith}, M. S.~M., {Coil}, A.~L., {et~al.} 2014, \apj, 784,
  128

\bibitem[{{Sobacchi} \& {Mesinger}(2015)}]{SobacchiMesinger2015}
{Sobacchi}, E., \& {Mesinger}, A. 2015, \mnras, 453, 1843

\bibitem[{{Sobral} {et~al.}(2018){Sobral}, {Santos}, {Matthee},
  {Paulino-Afonso}, {Ribeiro}, {Calhau}, \& {Khostovan}}]{Sobral2018a}
{Sobral}, D., {Santos}, S., {Matthee}, J., {et~al.} 2018, \mnras, 476, 4725

\bibitem[{{Sobral} {et~al.}(2017){Sobral}, {Matthee}, {Best}, {Stroe},
  {R{\"o}ttgering}, {Oteo}, {Smail}, {Morabito}, \&
  {Paulino-Afonso}}]{Sobral2017}
{Sobral}, D., {Matthee}, J., {Best}, P., {et~al.} 2017, \mnras, 466, 1242

\bibitem[{{Springel}(2005)}]{GADGET2_2005}
{Springel}, V. 2005, \mnras, 364, 1105

\bibitem[{{Szapudi}(1998)}]{Szapudi1998}
{Szapudi}, I. 1998, \apj, 497, 16

\bibitem[{{Taniguchi} {et~al.}(2005){Taniguchi}, {Ajiki}, {Nagao}, {Shioya},
  {Murayama}, {Kashikawa}, {Kodaira}, {Kaifu}, {Ando}, {Karoji}, {Akiyama},
  {Aoki}, {Doi}, {Fujita}, {Furusawa}, {Hayashino}, {Iwamuro}, {Iye},
  {Kobayashi}, {Kodama}, {Komiyama}, {Matsuda}, {Miyazaki}, {Mizumoto},
  {Morokuma}, {Motohara}, {Nariai}, {Ohta}, {Ohyama}, {Okamura}, {Ouchi},
  {Sasaki}, {Sato}, {Sekiguchi}, {Shimasaku}, {Tamura}, {Umemura}, {Yamada},
  {Yasuda}, \& {Yoshida}}]{Taniguchi2005}
{Taniguchi}, Y., {Ajiki}, M., {Nagao}, T., {et~al.} 2005, Publications of the
  Astronomical Society of Japan, 57, 165

\bibitem[{{Thacker} \& {Couchman}(2006)}]{LagrangePerturb_ThnC}
{Thacker}, R.~J., \& {Couchman}, H.~M.~P. 2006, Computer Physics
  Communications, 174, 540

\bibitem[{{Tilvi} {et~al.}(2009){Tilvi}, {Malhotra}, {Rhoads}, {Scannapieco},
  {Thacker}, {Iliev}, \& {Mellema}}]{Tilvi2009}
{Tilvi}, V., {Malhotra}, S., {Rhoads}, J.~E., {et~al.} 2009, \apj, 704, 724

\bibitem[{{Tilvi} {et~al.}(2014){Tilvi}, {Papovich}, {Finkelstein}, {Long},
  {Song}, {Dickinson}, {Ferguson}, {Koekemoer}, {Giavalisco}, \&
  {Mobasher}}]{Tilvi2014}
{Tilvi}, V., {Papovich}, C., {Finkelstein}, S.~L., {et~al.} 2014, \apj, 794, 5

\bibitem[{{Tinker} {et~al.}(2008){Tinker}, {Conroy}, {Norberg}, {Patiri},
  {Weinberg}, \& {Warren}}]{Tinker2008}
{Tinker}, J.~L., {Conroy}, C., {Norberg}, P., {et~al.} 2008, \apj, 686, 53

\bibitem[{{Tinker} {et~al.}(2006){Tinker}, {Weinberg}, \&
  {Warren}}]{Tinker2006}
{Tinker}, J.~L., {Weinberg}, D.~H., \& {Warren}, M.~S. 2006, \apj, 647, 737

\bibitem[{{Totsuji} \& {Kihara}(1969)}]{Totsuji1969}
{Totsuji}, H., \& {Kihara}, T. 1969, Publications of the Astronomical Society
  of Japan, 21, 221

\bibitem[{{Trainor} {et~al.}(2019){Trainor}, {Strom}, {Steidel}, {Rudie},
  {Chen}, \& {Theios}}]{Trainor2019}
{Trainor}, R.~F., {Strom}, A.~L., {Steidel}, C.~C., {et~al.} 2019, \apj, 887,
  85

\bibitem[{{Uhlemann} {et~al.}(2020){Uhlemann}, {Friedrich},
  {Villaescusa-Navarro}, {Banerjee}, \& {Codis}}]{Uhlemann2020}
{Uhlemann}, C., {Friedrich}, O., {Villaescusa-Navarro}, F., {Banerjee}, A., \&
  {Codis}, S.~r. 2020, \mnras, 495, 4006

\bibitem[{{Walsh} \& {Tinker}(2019)}]{WalshTinker2019}
{Walsh}, K., \& {Tinker}, J. 2019, \mnras, 488, 470

\bibitem[{{Wang} {et~al.}(2019){Wang}, {Mao}, {Zentner}, {van den Bosch},
  {Lange}, {Schafer}, {Villarreal}, {Hearin}, \& {Campbell}}]{Wang2019}
{Wang}, K., {Mao}, Y.-Y., {Zentner}, A.~R., {et~al.} 2019, \mnras, 488, 3541

\bibitem[{{Wen} {et~al.}(2020){Wen}, {Kemball}, \& {Saslaw}}]{Wen2020}
{Wen}, D., {Kemball}, A.~J., \& {Saslaw}, W.~C. 2020, arXiv e-prints,
  arXiv:2001.06119

\bibitem[{{White}(1979)}]{White1979}
{White}, S.~D.~M. 1979, \mnras, 186, 145

\bibitem[{{Wolk} {et~al.}(2013){Wolk}, {McCracken}, {Colombi}, {Fry},
  {Kilbinger}, {Hudelot}, {Mellier}, \& {Ilbert}}]{Wolk2013}
{Wolk}, M., {McCracken}, H.~J., {Colombi}, S., {et~al.} 2013, \mnras, 435, 2

\bibitem[{{Wright}(2006)}]{NedWright}
{Wright}, E.~L. 2006, \pasp, 118, 1711

\bibitem[{{Yang} \& {Saslaw}(2011)}]{YangSaslaw2011}
{Yang}, A., \& {Saslaw}, W.~C. 2011, \apj, 729, 123

\bibitem[{{Zehavi} {et~al.}(2002){Zehavi}, {Blanton}, {Frieman}, {Weinberg},
  {Mo}, {Strauss}, {Anderson}, {Annis}, {Bahcall}, {Bernardi}, {Briggs},
  {Brinkmann}, {Burles}, {Carey}, {Castander}, {Connolly}, {Csabai},
  {Dalcanton}, {Dodelson}, {Doi}, {Eisenstein}, {Evans}, {Finkbeiner},
  {Friedman}, {Fukugita}, {Gunn}, {Hennessy}, {Hindsley}, {Ivezi{\'c}}, {Kent},
  {Knapp}, {Kron}, {Kunszt}, {Lamb}, {Leger}, {Long}, {Loveday}, {Lupton},
  {McKay}, {Meiksin}, {Merrelli}, {Munn}, {Narayanan}, {Newcomb}, {Nichol},
  {Owen}, {Peoples}, {Pope}, {Rockosi}, {Schlegel}, {Schneider}, {Scoccimarro},
  {Sheth}, {Siegmund}, {Smee}, {Snir}, {Stebbins}, {Stoughton}, {SubbaRao},
  {Szalay}, {Szapudi}, {Tegmark}, {Tucker}, {Uomoto}, {Vanden Berk}, {Vogeley},
  {Waddell}, {Yanny}, \& {York}}]{Zehavi2002}
{Zehavi}, I., {Blanton}, M.~R., {Frieman}, J.~A., {et~al.} 2002, \apj, 571, 172

\bibitem[{{Zheng} {et~al.}(2010){Zheng}, {Cen}, {Trac}, \&
  {Miralda-Escud{\'e}}}]{Zheng2010}
{Zheng}, Z., {Cen}, R., {Trac}, H., \& {Miralda-Escud{\'e}}, J. 2010, \apj,
  716, 574

\bibitem[{{Zheng} {et~al.}(2016){Zheng}, {Malhotra}, {Rhoads}, {Finkelstein},
  {Wang}, {Jiang}, \& {Cai}}]{Zheng2016}
{Zheng}, Z.-Y., {Malhotra}, S., {Rhoads}, J.~E., {et~al.} 2016, The
  Astrophysical Journal Supplement Series, 226, 23

\end{thebibliography}

\end{document}